\documentclass[prb,twocolumn,floatfix]{revtex4}
\usepackage{graphics}
\usepackage{graphicx}
\usepackage{float}
\usepackage{amsfonts}
\usepackage{amssymb}
\usepackage{textcomp}
\usepackage{color}
\def\s3s3{$\sqrt{3}\times\sqrt{3}$}

\begin{document}
\title{Semi-classical spin dynamics of the antiferromagnetic
  Heisenberg model on the kagome lattice}

\author{Mathieu Taillefumier$^{1,2,3}$, Julien Robert$^{4}$,
  Christopher L. Henley$^{5}$, Roderich Moessner$^{6}$, and Benjamin
  Canals$^{7}$} \affiliation{$^1$Okinawa Institute of sciences and
  technologies, 1919-1 Tancha-son, Okinawa 904-0495 Japan}
\affiliation{$^2$J. W. Goethe Universit\"at, Institute f\"ur
  theoretische Physik, Max Von Laue Str.\ 1, 60438 Frankfurt-am-Main,
  Germany} \affiliation{$^3$UIO, Department of physics, P.O box 1048,
  Blindern 0316, Oslo Norway} \affiliation{$^4$Laboratoire L\'eon
  Brillouin CEA Saclay, b\^at.563 91191 Gif-sur-Yvette Cedex, France}
\affiliation{$^5$LASSP, Clark Hall, Cornell University, Ithaca, NY
  14853-2501, USA} \affiliation{$^6$Max Planck Institute for the
  complex systems, Noethnitzer Str. 38 D-01187 Dresden, Germany}
\affiliation{$^7$Institut N\'eel, CNRS/UJF, 25 avenue des Martyrs, BP
  166, 38042 Grenoble, Cedex 09, France}

\begin{abstract}
  We investigate the dynamical properties of the classical
  antiferromagnetic Heisenberg model on the kagome lattice using a
  combination of Monte Carlo and molecular dynamics simulations.
  We find that frustration induces a distribution of timescales in the
  cooperative paramagnetic regime ({\it i.e.} far above the onset of
  coplanarity), as recently reported experimentally in deuterium
  jarosite.
  At lower temperature, when the coplanar correlations are well
  established, we show that the weathervane loop fluctuations control
  the system relaxation : the time distribution observed at higher
  temperatures splits into two distinct timescales associated with
  fluctuations in the plane and out of the plane of coplanarity. The
  temperature and wave vector dependences of these two components are
  qualitatively consistent with loops diffusing in the entropically
  dominated free energy landscape.
  Numerical results are discussed and compared with the $O(N)$ model
  and recent experiments for both classical and quantum realizations
  of the kagome magnets.
\end{abstract}

\maketitle
\section{Introduction}

In psychology, frustration is an emotional response to opposition or
conflict. In the natural sciences, frustration is often associated
with the impossibility of a system to optimize simultaneously all
elementary interactions, whether they are single body, two body, or
many body. It gives rise to many exciting phenomena in particular in
magnets where magnetic states stay disordered despite the presence of
strong interactions.

Although the first studies of spins models with competing interactions
date back from the early 50s~\cite{Wannier1950,Kano1953}, the
terminology of frustration was introduced in the 70s mostly in the
context of spin
glasses\cite{Toulouse1977,Kirk1977,Kar1979,Villain1979}. Since then it
has been associated to many unconventional low energy states such as
quantum and classical spin
liquids\cite{Anderson1973,Reimers1993,Moessner1998,Balents2010,Misguich2008,Hermele2008}, or
more recently, to quantum, classical and artificial spin
ices\cite{Harris1997,Molavian2007,Nisoli2013,Gingras2011}.

All these phenomena take place in diverse systems but each of them can
be associated with a canonical representative, {\it i.e.} a minimal
frustrated spin model which brings together most of the important
features.

The 2D antiferromagnetic Heisenberg model on the kagome lattice
(KHAFM) is one of the archetypes of such systems.  The kagome lattice
can be described as a lattice of triangles sharing one corner with
each neighbor, the key property for creating these unusual and often
highly degenerate ground states both in classical and quantum
models\cite{Huse1992,Chalker1992,Sachdev1992,Reimers1993,Harris1997,Molavian2007}.

In the quantum limit, the ground state of KHAFM is still unsettled,
but recent results rather point towards a quantum disordered ground
state\cite{Yan2011,Messio2012,Depenbrock2012,Iqbal2013}. In the
classical limit, the equilibrium properties of KHAFM in the
paramagnetic ($T>|J|$) and cooperative regimes ($0.001 |J| < T < |J|$)
regimes are now well known and understood down to low
temperature\cite{Huse1992,Chalker1992,Reimers1992,Reimers1993,Harris1992,Zhitomirsky2008,Henley2009,Cepas2012}
(where $|J|$ is the nearest neighbor interaction coupling constant).
Finally a recent study finds a very weak magnetic order when the system is
deep in the coplanar regime ($T<0.001 |J|$)\cite{Chern2013}.

The dynamics of the classical Heisenberg model in both one, two or
three dimensions is very rich at all temperatures because of the non
linearity of the model. The dynamics in the paramagnetic regime was
extensively studied during the nineties. 

Early studies of the dynamics of frustrated magnets have shown that
these systems are very different from their non-frustrated
counterpart\cite{Keren1994} at low temperature. In classical 3D
frustrated magnets, such as pyrochlore antiferromagnets, most
correlations in magnetic states decay exponentially at low temperature
while the temperature dependence of the relaxation time follows a
power law\cite{Moessner1998,Conlon2009,Gardner2010}. Similar results
were found in the kagome antiferromagnets despite the more complex
landscape around the ground state
manifold\cite{Robert2008,Schnabel2012}.

At low temperature, the absence of long range ordered ground states
does not forbid short-lived spinwave like
excitations\cite{Halperin1977} whose natural time scale is of the
order of $|J|^{-1}$. The highly degenerate nature of the ground state
manifold gives rise to additional processes that contribute at
intermediate time
scales\cite{Moessner1998,Robert2008,Schnabel2012}. The longest time
scales ($t |J| > 500$) are the domain of both (i) spin diffusion, {\it
  i.e.} the stochastic propagation of the magnetization throughout
large magnetic regions, and of (ii) magnetic relaxation, {\it i.e.}
the gradual reorganization of the average spin configuration around
which the spin waves are oscillating. All these time scales are
present in KHAFM and can be studied with different experimental
probes, ranging from thermal neutron scattering and muon relaxation
for the short and intermediate time scales to ac-susceptibility and
NMR measurements for the long time scales.

The aim of this work is to understand and characterize the dynamics
beyond short time scales. We find that spin diffusion persists at low
temperature ($T/|J| < 10 ^{-2}$) despite the presence of strong spin
correlations. Below $T/|J| = 10 ^{-2}$, the spin dynamics becomes
anisotropic due to the entropic selection of coplanar states. We also
find that the relaxation is mediated by large amplitude oscillations
around small loops (also called weathervanes defects) and
spinwaves, despite the absence of long range order.

\begin{figure*}[!t]
  \centering
  \includegraphics[width=0.8\textwidth]{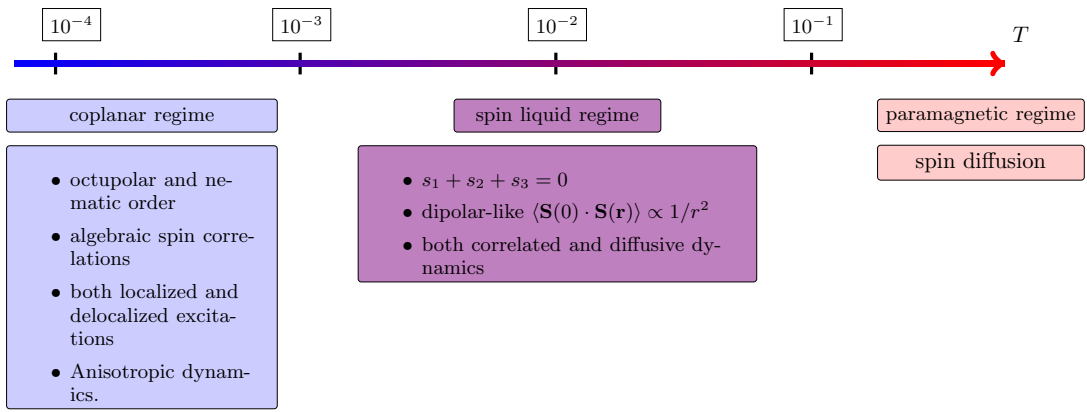}
  \caption{(color online) Schematic phase diagram of the classical
    Heisenberg model on the kagome lattice. This model undergoes two
    different crossovers when the temperature decreases. The
    paramagnetic regime has conventional diffusive behaviors in the
    high $T$ regime. The first crossover occurs around $T\approx J$
    and the system enters a regime where spin correlations develop at
    a length scale of the order of the spin correlation length
    $\xi(T)$. Around $T\approx 10^{-3} J$, the entropic selection
    favors coplanar states whose dynamics is anisotropic.}
\label{fig:diagram}
\end{figure*}

This article is organized as follows.
In section \ref{sec:thermodynamics}, current knowledge of KHAFM
thermal behavior is reviewed in order to provide the reader with a
clear description of the magnetic structures that the spin dynamics is built
on. Then, numerical procedures and technical details are given in
section \ref{sec:model}. Finally, our numerical results are presented
in sections \ref{sec:paraliquid} and \ref{ssec:coplanar} respectively
for the high and low temperature regimes, and compared with recent
experimental results obtained in kagome systems (see
Sec. \ref{sec:exp}).

\section{Equilibrium properties of the classical kagome Heisenberg
  antiferromagnet - foundations for a spin dynamics}
 \label{sec:thermodynamics}

 The phase diagram of the classical KHAFM is depicted on
 Fig.~\ref{fig:diagram}.  At high temperatures ($T > |J|$), the
 classical KHAFM is a conventional paramagnet with short range
 spin-spin correlations.
 When temperature becomes comparable to the exchange $|J|$,
 correlations appear and spins on each triangular plaquette of the
 kagome lattice approximately sum to zero and are oriented at
 120\textdegree \ to one another.
 This local arrangement does not lead to large correlated domains
 because of the excitence of an uncountable number of configurations
 that form a highly degenerate and connected manifold associated with
 ``origami'' folding of the spin
 pattern~\cite{Ritchey1993,Shender1993,Shender1995}. As a result of
 this degeneracy, spin correlations decay algebraically with distance
 and can be associated with a so-called Coulomb
 phase~\cite{Huse1992,Moessner1998,Henley2001,Isakov2004,Henley2009,Sen2012}.
 In such a phase, correlations are expected to decay algebraically
 with distance, with geometrical factors that depend on the chosen
 direction in the kagome lattice.
 This regime roughly covers the temperature range $5.10^{-3} < T/J <
 1$.

 When temperature is further reduced, {\it i.e} $T/J < 5.10^{-3}$, the
 free energy of all spin foldings is no longer uniform and the spins,
 which are still locally constrained to stay at 120\textdegree \
 within each triangle, now select a particular spin plane\footnote{By
   the Mermin Wagner theorem, the corresponding orders are cut-off at
   the longest lengthscales, leaving well-defined regimes separated by
   cross-overs.}, common over many triangles, around which they are
 fluctuating~\cite{Chalker1992}.

 This selection of coplanar states, also known as entropic ordering
 (or order out of disorder), is due to the additional soft degrees of
 freedom for the thermal
 fluctuations~\cite{Moessner1998,Chalker1992,Ritchey1993,Chandra1993}
 available in the coplanar states. This was first identified as a
 coplanar ordering, {\it i.e.} the development of quadrupolar (or spin
 nematic) correlations\cite{Chalker1992}. This incipient order is not
 merely coplanar but was later recognized to imply octupolar order as
 well\cite{Zhitomirsky2008}.

 Thus, in the model's ultra low temperature regime, spins fluctuate
 around one of the discrete coplanar ground states, in which every
 spin has one of three possible directions, which can be represented
 by the values (or colors) of the discrete spins in the $3$-state
 Potts model on the same lattice.  The coplanar ground states
 correspond 1-to-1 up to global rotations to Potts ground states, in
 which every triangle has three colors
 \cite{Huse1992,Shender1995,Henley2009}, whose number is $N_c \approx
 1.13^N$ where $N$ is the number of spins of the lattice (we will
 always consider finite lattices with periodic boundary conditions).

 Consequently, there are essentially three different regimes.  The
 generic paramagnetic regime with short range spin correlations, a
 cooperative paramagnetic regime or spin liquid regime, with algebraic
 correlations on finite area domains, whose area is controlled by a
 temperature dependent correlation length and a nematic-like regime,
 where correlations are enhanced via an order out of disorder
 phenomenon that stabilizes a common spin plane.  At very low
 temperature magnetic ordering also appears~[\onlinecite{Chern2013}].

 The first studies of the dynamics of magnetic systems concentrated on
 the nature of spin fluctuations in the cooperative paramagnetic
 regime in comparing the spin dynamics of a strongly correlated
 disordered magnet with the dynamics of an ordered one.
 Following this perspective, it was shown that the KHAFM is a model
 with unusually high density of low lying excitation\cite{Keren1994}
 at low temperatures.
 It was also shown that at sufficiently low temperatures ($T/J \le
 5.10^{-3} $), coherent excitations are unexpectedly stable despite
 being built on a thermodynamically characterized disordered
 manifold\cite{Robert2008}.

 In this work, our interest is to understand how the natural high
 temperature ($T \gg J$) signature of diffusive dynamics is found at
 lower temperatures ($T \ll J$), how it terminates close to the
 nematic boundary even though local excitation are still present (6-sites
 loops), and how this model discriminates between in-plane and
 out-of-plane spin dynamics, all latter considerations being discussed
 in the intermediate $10 \, J^{-1} \le \tau \le 10^4 \, J^{-1}$ time
 scale.
 In other words, it aims at resolving spin fluctuations 
 whilst extending previous dynamical studies in order to cover a time
 range associated with magnetic relaxation rather than structured and
 propagative excitation.

\section{Model, numerical procedures and overview of the results}
\label{sec:model}

In this section, we first define the model and the notation we use in
this manuscript, as well as the method we use to investigate the spin
dynamics at finite temperature.

The numerical procedures used to perform the stochastic sampling of the
phase space and to integrate the non linear equations of motions are
then detailed.
Based on this technical framework, we justify our choice of
temperature range and lattice sizes to ensure that most of the
discussed results are free of finite size effects.

We end this section with a short overview of the dynamics in the three
temperature regimes that will be developed in the following sections.

\subsection{Model}

We consider the classical Heisenberg model
\begin{equation}
  \label{eq:m:1}
  \mathcal{H} = - J \sum_{\left<i,j\right>} {\bf s}_i\cdot {\bf s}_j,
\end{equation}
where the summation is limited to nearest neighbors, $J<0$ is the
isotropic antiferromagnetic coupling constant and $|{\bf s}_i|=1$ are
classical spins on the unit sphere $\mathcal{S}^2$ located at the
kagome sites.

The kagome lattice is described as a two dimensional triangular
lattice with a triangular unit cell and displacement vectors ${\bf a}
= a(1,0)$ and ${\bf b}=a(-1/2,\sqrt{3}/2)$, with $a$ the lattice
constant. The unit cell contains three spins at positions ${\bf
  r}_1=(0,0)$, ${\bf r}_2={\bf a}/2$ and ${\bf r}_3={\bf b}/2$. The
index $i=({\bf R}_i,\alpha_i)$ in Eq.~(\ref{eq:m:1}) is a compact
notation that regroups both the position ${\bf R}_i$ of the unit cell
where the spin resides and $\alpha_i$ its sublattice index. With these
notations, the Brillouin Zone (BZ) is an hexagon with corners located
at $(Q_{a},Q_{b})=\pm(1/3,1/3)$, $\pm(2/3,-1/3)$, $\pm(1/3,-2/3)$ in
reciprocal space with $(Q_{a},Q_{b})=Q_{a}\mathbf{a}^\star +
Q_{b}\mathbf{b}^\star$.

It is convenient to express Eq.~(\ref{eq:m:1}) as
\begin{equation}
  \label{eq:m:3}
  H = - \frac{J}{2} \sum_\eta {\bf l}^2_\eta + E_0,
\end{equation}
where $E_0$ is a constant energy shift and ${\bf l}_\eta=\sum_{i\in
  \eta} \mathbf{s}_i$ is the total spin of triangle $\eta$. From this
expression, it is possible to see that the ground state satisfies ${\bf
  l}_\eta = 0$ for all triangles, thus leading to a relative angle of
$\pm2\pi/3$ between neighboring spins in any ground state.

In this article, our interest lies in the time evolution of the
spin-pair correlations emerging in such a model. It is convenient to
probe such dynamical correlations in reciprocal space by calculating
the scattering function, also called dynamical structure factor 
\begin{equation}
S({\bf Q},t) = \langle {\bf s}_{\bf -Q}(0) \cdot {\bf s}_{\bf Q}(t) \rangle,
\label{eq:m:5}
\end{equation}
with
\begin{equation}
{\bf s}_{\bf Q}(t) = \sum_{i,\alpha} {\bf s}_{i,\alpha}(t) \mathrm{e}^{-i ({\bf R}_i +{\bf r}_\alpha)\cdot {\bf Q}}
\label{eq:m:6}
\end{equation}
${\bf R}_i$ and ${\bf r}_\alpha$ are respectively the position of the
unit cell and the coordinates of sublattice $\alpha$.

In expression (\ref{eq:m:5}), the semi classical spin dynamics at
$T=0$ is described by the non-linear Bloch equations
\begin{equation}
\frac{d{\bf s}_i(t)}{d t} = - J {\bf s}_i(t) \times \left(\sum_{j} {\bf s}_{j}(t)\right),
\label{eq:m:4}
\end{equation}
where sites $j$ are the nearest neighbors of $i$. Note that the set
described by Eq.~(\ref{eq:m:4}) conserves the total energy $E_{tot}$
and magnetization $M_{tot}$.
\begin{figure}[!ht]
\centering
\includegraphics[width=6cm]{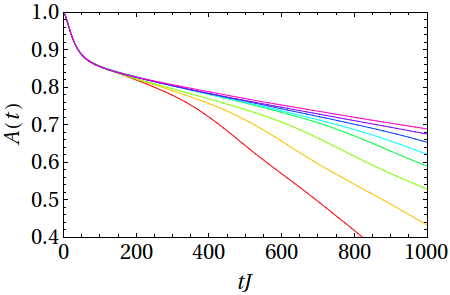}\\
\includegraphics[width=6cm]{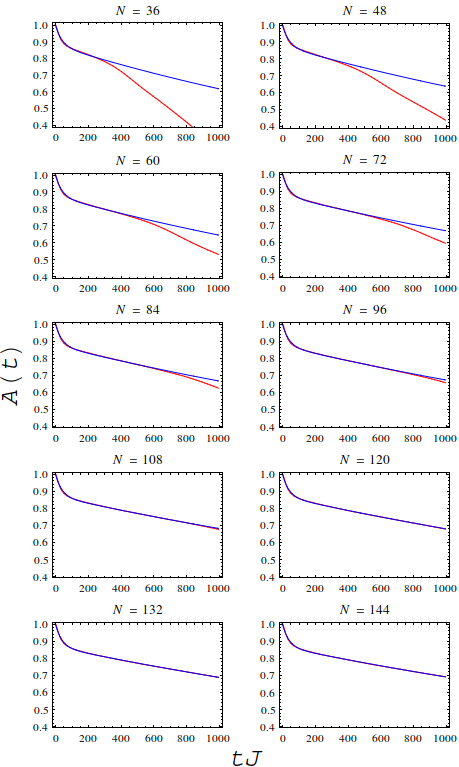}\\
\includegraphics[width=6cm]{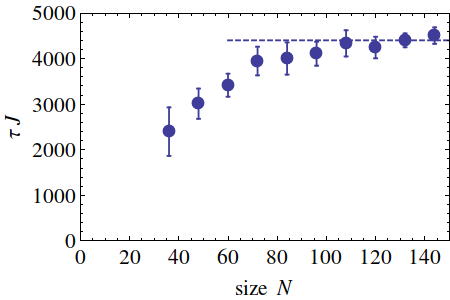}
\caption{(Color online) Finite size effects on the autocorrelation
  function $A(t)$ and relaxation time in the octupolar regime. (a)
  Autocorrelation function $A(t)$ at $T/J= 0.0006$ for different
  lattice size from $L=36$ (red) to $L=144$ (blue). (b) Fit (blue) of
  the numerical data (red) for the different lattice sizes ranging
  from $L=36$ to $L=144$, assuming that $A(t)=a_\parallel e^{-
    t/\tau_\parallel} + a_\perp e^{- t/\tau_\perp}$. (c) relaxation
  time $\tau_\parallel$ versus lattice size resulting from the fit
  shown in (b).}
\label{fig:finitesize}
\end{figure}

we numerically integrate Eq.~(\ref{eq:m:4}) in order to evaluate the
scattering function. We combine the deterministic integration of the
equations of motion with (hybrid) Monte Carlo simulations for
generating samples of spin arrays at a given temperature. This
numerical procedure is detailed in the next section.

\subsection{Numerical procedures}

The numerical integration of Eq.~(\ref{eq:m:4}) has been performed up
to 1024~$J^{-1}$ (even up to 10$^4$~$J^{-1}$ in the coplanar regime,
see section \ref{ssec:gen-biblio-predictions}) using an 8th-order
explicit Runge-Kutta (RK) method with an adaptative step-size control,
offering an excellent compromise between accuracy and computation
time.
The RK error parameter as well as the RK order have been fixed in
order to preserve the Euclidean distance $d=[\sum_i ({\bf s}_i^{RK} -
{\bf s}_i^{BS})^2]^{1/2}$, {\it i.e.} the distance between time
trajectories obtained with the RK method and with the more robust but
time consuming Burlisch-Stoer (BS) algorithm\cite{Hairer}.
As a result, trivial constants of motion, such as the total energy
$E_{\textrm{tot}}$ and magnetization $M_{\textrm{tot}}$, are conserved
with a relative error smaller than $10^{-9}$.

The initial spin configurations used for the numerical integration are
generated at each temperature by an hybrid Monte Carlo method using a
single spin-flip Metropolis algorithm. The single spin flip algorithm
becomes inefficient at low temperature because the number of rejected
attempts increases due to the development of spin correlation as the
system enters the liquid and the spin nematic regimes. To partially
overcome this effect, we reduce the solid angle from which each spin
flip trial is taken to ensure that the acceptance rate is above $0.4$
at every temperature.

Thousand spin configurations are used at each temperature to
evaluate the ensemble average in Eq.~(\ref{eq:m:5}) while the number
of Monte Carlo steps needed for decorrelation is adapted in such a way
that the stochastic correlation between spin configurations is lower
than $0.1$.

In the coplanar regime, the stochastic correlation between spin
configurations is relatively high because the number of accepted 
attempts is small. The system is trapped in the immediate surroundings
of one given coplanar configuration which means that ensemble
averaging is only representative of the initial conditions. To limit
this effect, we use an hybrid Monte Carlo Metropolis algorithm that
combines both over-relaxation\cite{Creuz1987} and the molecular
dynamics method described earlier.

These two methods correspond to rather different ways of exploring the
configuration space. An over-relaxation move, which fulfills the
detailed balance, consists of rotations the selected spin by a random
angle around its local exchange field so the system does not remain
precisely in the same spin configuration when the single spin flip is
rejected. However it does not prevent the system from being trapped
into the immediate surroundings of one given coplanar configuration,
so a huge number of Monte Carlo steps are still necessary for the
system to decorrelate.

On the other hand, as shown in this paper, the molecular dynamics
procedure is a very efficient way to probe different coloring
states (or Potts states) related  to each other by a spatially
localized excitation. Indeed, two-color closed spin loops of small
size are easily flipped while integrating the equation of motions,
even at temperatures as low as $T/J=0.0001$. Thus, our method acts as
a ``natural'' loop algorithm although the method is limited to small
loops as the flipping time grows rapidly with loop size and
temperature.

The numerical results have been obtained for different lattice sizes
ranging from $L=144$ (for the $\mathbf{Q}$-resolved scattering
function $S(\mathbf{Q},t)$) to $L=192$ (for the autocorrelation function
$A(t)$) with periodic boundary conditions, so the total number of
spins does not exceed $N=3L^2\lesssim 1.2\, 10^5$.
%
%

Finite size effects, which are negligible at high temperature, become
particularly important at low
temperature. Fig.\ref{fig:finitesize}-(a) shows that the evolution of
the autocorrelation function
\begin{eqnarray}
  A(t)& = & \int d^2\mathbf{Q}\, S(\mathbf{Q},t)\\
  & = & \sum_i \langle S_i(0) \cdot S_i(t) \rangle
\label{eq:at}
\end{eqnarray}
at $T/J=0.0006$ for different lattice sizes from $L=36$ (red) to $L=144$
(blue) becomes nearly independent of the system size when $L >
108$. Moreover the long time dynamics is affected by the rotation of
the spins around the residual magnetization~[\onlinecite{Moessner1998}]
\footnote{This effect is relatively small for large systems
  and can be compensated using the properties of the octupolar phase
  without modifying the equation of motions. It is worth noting that no
  finite size effects has been observed for the short time
  scales. Therefore, the analysis of the spin wave spectrum determined
  in ref. [\onlinecite{Robert2008}], as well as the evolution of their
  lifetime versus temperature is justified.
  However, the long time dynamics qualitatively discussed in
  ref. [\onlinecite{Robert2008}] (see Fig. 3 (b)) is affected by
  finite size effects in the coplanar regime.}.

The fit of the autocorrelation is represented for each lattice size on
Fig. \ref{fig:finitesize} (b) (for the fitting process, see section
\ref{ssec:numcop} and Eq. (\ref{eq:Atparaperp})). While the short time
relaxation ($tJ<60$) does not depend on the lattice size, the long
time relaxation, plotted versus $L$ in Fig. \ref{fig:finitesize} (c),
does not seem to vary significantly for $L>120$.
Consequently, finite size effects will be neglected in the following
for $T/J \gtrsim 0.0001$ and $L\geq 144$. Microscopic quantities like
the correlation lengths or more generally the $\mathbf{Q}$-resolved
scattering function $S(\mathbf{Q},t)$ may however be affected by
finite size effect at least in certain regions of reciprocal
space. This problem will be discussed in section \ref{ssec:coplanar}.

\subsection{Three dynamical regimes with blurred boundaries}
\begin{figure}[!ht]
\includegraphics[width=7.5cm]{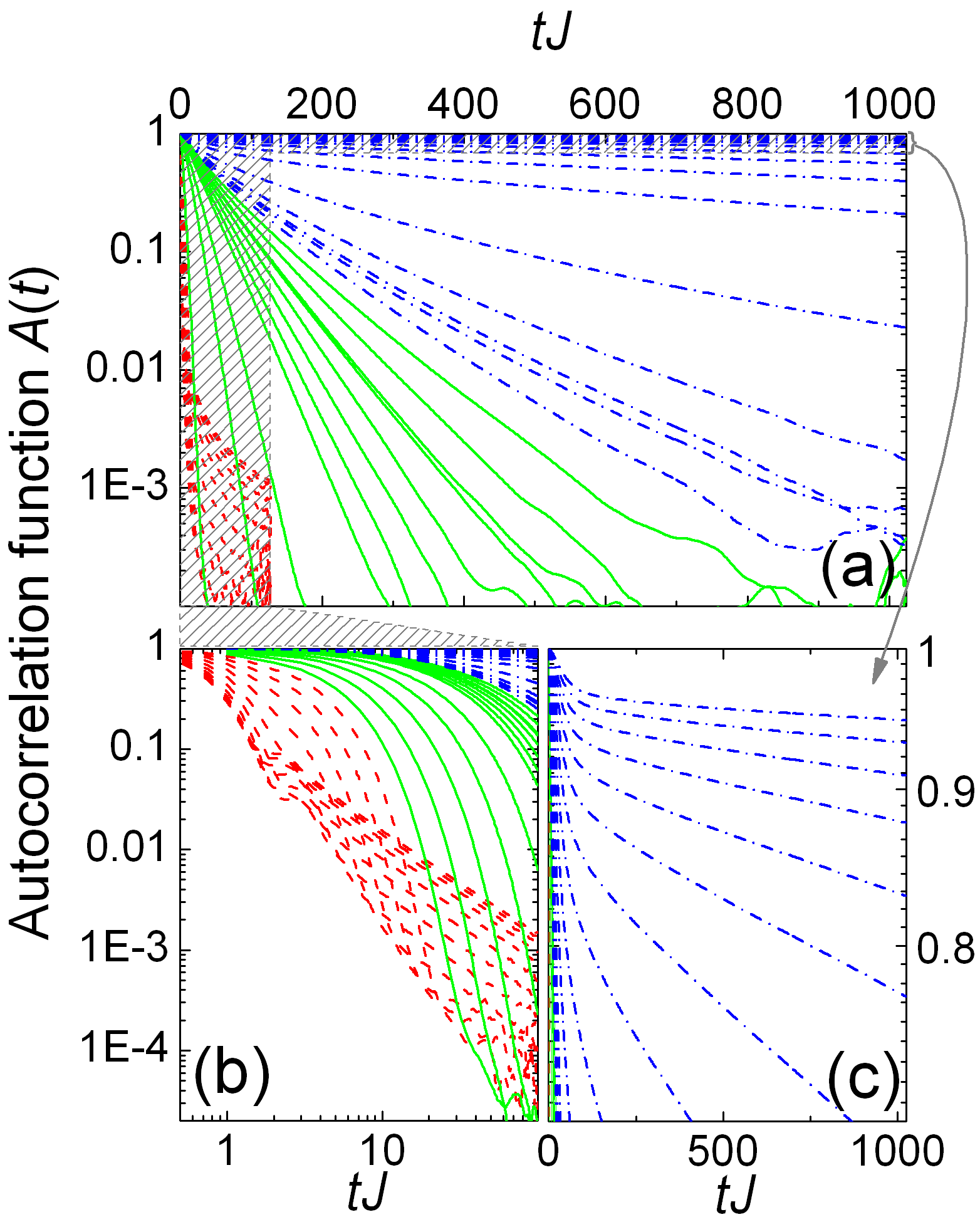}
\caption{(Color online) Temperature dependence of the semi-classical
  spin dynamics as releaved by the autocorrelation function $A(t)$.
  Autocorrelation function $A(t)$ versus time in the paramagnetic
  ($T/J=10-0.1$), cooperative ($T/J=0.1-0.005$) and coplanar
  ($T/J<0.005$) regimes respectively represented by dashed-red,
  solid-green and dotted-dashed-blue lines: (a) linear-log scale, (b)
  log-log scale, and (c) linear-log scale focusing on the coplanar
  regime.}
\label{fig:autocorr}
\end{figure}

Our main goal in this work is to probe the fluctuations around the
ground state manifold.
Before characterizing the relaxation dynamics and establishing in
particular the temperature and wave-vector dependence of the lifetime
of the correlated magnetic domains, we qualitatively discuss the
temperature dependence of a global quantity, the autocorrelation
function $A(t)$ defined by Eq.~(\ref{eq:at}).
A more detailed study of each regime is given in sections
\ref{sec:paraliquid} and \ref{ssec:coplanar}.

The autocorrelation function $A(t)$, shown in Fig. \ref{fig:autocorr}
(a)-(c), is respectively represented by dashed-red, solid-green and
dotted-dashed-blue lines for the paramagnetic, spin liquid and
coplanar regimes. Linear-log (a,c) and log-log (b) scales are used to
exhibit both exponential relaxations and diffusive behaviors. While it
is manifest from panel (a) that a slowing down of the spin
fluctuations with decreasing $T$ is at work - as could be expected for
any conventional magnetic system - we also notice that the overall
shape of $A(t)$ strongly depends on the temperature range.

In the paramagnetic regime, although $A(t) = 1-t^2\alpha$ for shortest
times, the linear tail in a log-log scale above $tJ \simeq 4$ (see
Fig.~\ref{fig:autocorr} (b)) is characteristic of spin diffusion
expected to be found in the limit of high temperatures and long
wave-lengths and times\cite{Muller1988,Gerling1989}. The signature of 
diffusive behavior is strongly reduced with decreasing temperature and
is no longer visible in the two lowest temperature
regimes. Nevertheless, it will be shown in section \ref{ssec:paraliqu}
that $(i)$ spin diffusion is still present in a slightly reduced
$q$-range with the onset of short-range correlations below $T/J=0.1$,
and $(ii)$ this range tends to zero at the octupolar transition (at
least, it becomes smaller than the wave-vector resolution, so that
there is no apparent diffusive behavior in our simulations for this
lattice size).

Below the paramagnetic/cooperative crossover occurring around $T/J\sim
0.1$, the rough linear dependence of $A(t)$ in a linear-log scale (see
Fig.~\ref{fig:autocorr}(a)) suggests an exponential decay
$\mathrm{e}^{-t/\tau_T}$ with a temperature dependent relaxation time
$\tau_T$. Nonetheless, the detailed analysis of $S(\mathbf{Q},t)$
given in the next section will highlight that $\tau_T$ is
$\mathbf{Q}$-dependent as well, so that only an average
appears in $A(t)$.

Finally, the most surprising feature in Fig. \ref{fig:autocorr} is
probably the intriguing behavior of $A(t)$ in the octupolar regime,
showing a kink in the $A(t)$ behavior at around $tJ \sim 60$. It is
related to the presence of two relaxation processes that are different
in nature (see sec.~\ref{ssec:coplanar} for more details).
In particular, it will be shown that the entropic selection $(i)$
strongly affects the fluctuations of the groundstate manifold far
above the transition toward coplanarity ($T/J\lesssim 0.05$, see
section \ref{ssec:paraliqu}), and $(ii)$ leads to different dynamical
behavior for the in-plane and out-of-plane spin components below the
transition (see sec.\ref{ssec:coplanar}).

\section{Paramagnetic and cooperative regimes}
\label{sec:paraliquid}

\subsection{Models, predictions}
\label{sec:diffusion}

In the absence of any order, the most basic dynamical process that may
happen in a simple Heisenberg spin model is a stochastic process
transferring spin density from a magnetic site to a neighboring
one. By a succession of such thermally-activated random steps, the
spin density arrives at a large distance ${\bf r}$ with a probability
given by phenomenological spin-diffusion
theory\cite{Marshall1968,Halperin1969,Muller1988}. Since the total magnetization is a
conserved quantity, the magnetization density ${\bf m}({\bf r},t)$
must fulfill a local equation of continuity
\begin{equation}
  \label{eq:a:1}
  \frac{\partial {\bf m}({\bf r},t)}{\partial t} + {\bf\nabla}.{\bf j}({\bf r},t) = 0.
\end{equation}
If we assume that the local current ${\bf j}({\bf r},t)$ is related
to the magnetization by Fick's first law
\begin{equation}
  \label{eq:a:2}
  {\bf j}({\bf r},t) = - D \nabla {\bf m}({\bf r},t),
\end{equation}
where $D$ is the diffusion coefficient that depends on the details of
the model, and after expressing Eq.~(\ref{eq:a:1}) and (\ref{eq:a:2})
in Fourier space, the magnetization density obeys the
diffusion equation
\begin{equation}
  \label{eq:a:3}
  \frac{\partial {\bf m}({\bf q},t)}{\partial t} = - D q^2 {\bf m}({\bf q}, t)
\end{equation}
in the hydrodynamic regime, {\it i.e.} for large time $t$ and
wave vectors smaller than the inverse of the correlation length
$q<1/\xi$ \cite{Halperin1969}. After integration over time of
Eq.\ref{eq:a:3}, 
\begin{equation}
  \label{eq:a:4}
  S({\bf q},t) = S({\bf q},0) \mathrm{e}^{-D q^2 t}.
\end{equation}
Integrating over $q$ gives rise to an autocorrelation with a tail that
follows a power law $A(t)\simeq t^{-d/2}$ where $d$ is the dimension
of the system\cite{Muller1988}.

At lower temperature $T\ll J$, the spin dynamics becomes sensitive to
the magnetic correlations which extend over scales of the order of
the spin correlation length $\xi$, which diverges as $1/T$ according to
the predictions for $N$-component spin model\cite{Garanin1999} (ICSM).
This model describes very well the apparition of structured spin pair
correlations in classical Heisenberg
systems\cite{Isakov2004,Henley2005}. Coupled to an appropriate
Langevin dynamics, it becomes a powerful method to predict the
temperature dependence of the dynamical properties\cite{Conlon2009}.

In this model\cite{Conlon2009} which we describe here for completeness, each
spin component in the large-$N$ limit has the normalized probability
distribution $e^{-\beta E}$ with
\begin{equation}
  \beta E = \frac12\sum_i \lambda s^2_i+\frac12 \beta J \sum_\alpha {\bf l}_\alpha^2.
  \label{eq:l:1}
\end{equation}
${\bf l}_\alpha = \sum_{i\in \alpha} s_i$ is the sum of the soft spins
$-\infty < s_i < \infty$ forming the triangle $\alpha$. The energy
function (\ref{eq:l:1}) differs from Eq.~(\ref{eq:m:4}) by an
additional term that constrains the length of the spins. The Lagrange
multiplier \mbox{$\lambda = 1 + O(T/J)$} in the limit $T\ll J$ is
obtained by imposing $\left<s^2_i\right>=1/3$ to each single component
of the spin to mimic the behavior of Heisenberg
spins\cite{Conlon2009}.

Then, the diffusive dynamics emerging from these static correlations
can be described by the Langevin equation
\begin{equation}
  \label{eq:l:2}
  \frac{d s_i}{dt} = \Gamma\sum_l \Delta_{il} \frac{\partial E}{\partial s_l} + \xi_i(t)
\end{equation}
for each spin component, and whose integration yields an
analytic expression of the dynamical scattering function
$S(\mathbf{q},t)$. In this expression, $\Delta_{ij} =
A^{\text{ad}}_{ij} - z \delta_{ij}$ is the lattice Laplacian, $z$ is
the coordination number of the lattice ($z=4$ for the kagome lattice)
and $A^{\text{ad}}_{ij}$ is the adjacency matrix (see the appendix for
details). $\Gamma$, which sets the energy scale of the dynamical
processes, is the only free parameter of the model.
This model contains two terms, a drift current that we take
proportional to the difference of generalized forces $\partial
E/\partial s_j$ on each bond of the lattice, and a second term
imposing thermal equilibrium described by a Gaussian noise
contribution $\xi_i(t)$ on each site $i$ of the lattice bonds. The
noise term is correlated with an amplitude fixed by the requirement of
thermal equilibrium:
\begin{equation}
  \label{eq:l:3}
  \left<\xi_i(t)\xi_j(t^\prime)\right> = 2 T \Gamma \Delta_{ij} \delta(t-t^\prime).
\end{equation}

This model was initially proposed for studying the dynamics of the
pyrochlore antiferromagnet in the limit $T\ll
J$\cite{Conlon2009}. Similar results are found for the KHAFM: around
the center of the Brillouin zone where the scattering function is described
by Eq.~(\ref{eq:a:4}) with a temperature independent diffusion
coefficient. At larger wavevectors and away from the high symmetry
directions, the decay rate $\tau_\alpha^{-1} \propto T$ varies
linearly with temperature and does not depend on $q$.

\subsection{Numerical results and discussion}
\label{ssec:paraliqu}

In the following subsections, we show that spin diffusion is observed
in the hydrodynamic regime as predicted in the previous section with,
however, a temperature dependent diffusion coefficient $D_T$. At
larger wave-vectors and away from the nodal lines $[h,0]$, $[0,h]$ and
$[h,-h]$, where the dynamical properties are dominated by
finite-energy spin wave-like excitation
\cite{Robert2008,Schnabel2012}, an exponential relaxation is observed
with a temperature and wave-vector dependent relaxation time
$\tau_T(\mathbf{Q})$ (sec. \ref{sssec:relaxmax}), revealing the
sizable effect of the entropic selection even at temperatures far
above the transition toward coplanarity.

\subsubsection{Hydrodynamic regime}
\label{sssec:hydro}
\begin{figure}[!t]
\includegraphics[width=8cm]{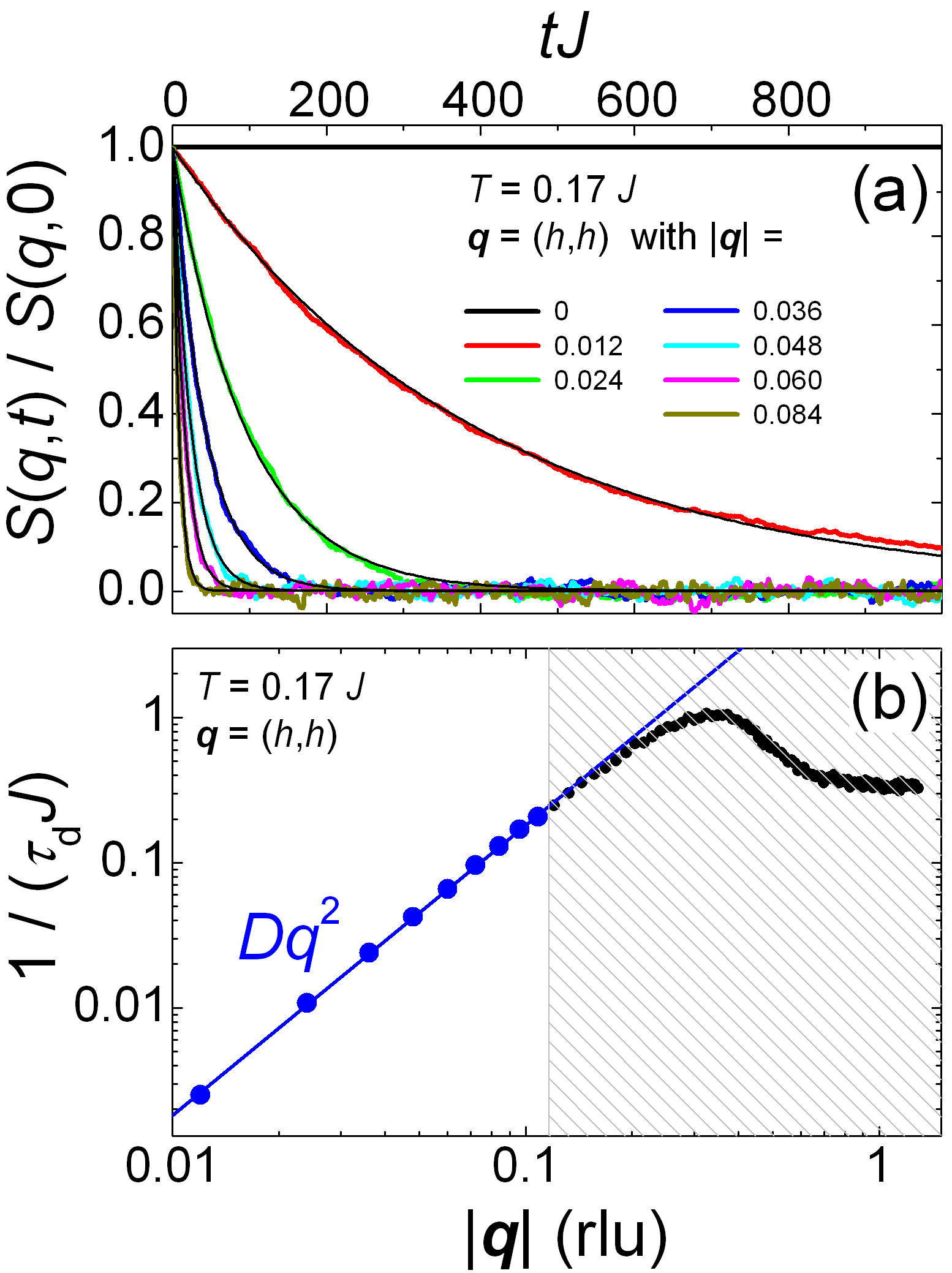}
\caption{(Color online) Spin diffusion in the cooperative regime as revealed
  by the spin-spin correlation functions. (a) Scattering function
  $S(q,t)/S(q,0)$ versus $tJ$ at $T/J=0.17$ for several wave-vectors
  close to the zone center in the $[h,h]$ direction. The fits, 
  performed at each wave-vector using a decaying exponential law (see
  text), are represented by the thin black lines. (b) Inverse of the
  relaxation time $\tau_d$ extracted from figure (a) as function of
  the wave-vector. The wave-vector region ($|q| \gtrsim 0.1$) where
  the spin diffusion law is not valid is shaded.}
\label{fig:diffusion1}
\end{figure}

In the hydrodynamic regime, expected to characterize only the
long-wavelength-low-frequency response of the system (see
sec.\ref{sec:diffusion}), the scattering function $S(\mathbf{Q},t)$ is
expected to decrease exponentially with a relaxation rate
$\tau^{-1}_T(q) = D_T q^2$ where $D_T$ is the diffusion coefficient
(see eq.(\ref{eq:a:4})). On the other hand, a short time expansion of
the scattering function leads to the following form\cite{Marshall1968}
\begin{eqnarray}
  \frac{S(\mathbf{Q},t)}{S(\mathbf{Q},0)} & = & 1 - \frac{1}{2}\langle \omega^2 \rangle t^2 + \frac{1}{24}\langle \omega^4 \rangle t^4 + \mathcal{O}(t^6),
\end{eqnarray}
with $\langle \omega^n \rangle = \int_{-\infty}^\infty{\omega^n
  S(\mathbf{Q},\omega)d\omega} / \int_{-\infty}^\infty
S(\mathbf{Q},\omega)d\omega$ the $n$-th moment of the normalized
spectral weight function. As for the $1D$ case\cite{Marshall1968}, we
find that the expansion up to the fourth order describes well the
numerical simulations for $0<tJ<1$. Above the spin velocity
correlation time $t_{vc} = \langle\omega^2 \rangle ^{-1/2}$ which is
no more than a few $J^{-1}$ at most wavevectors, the hydrodynamic
regime appears and diffusion occurs.
\begin{figure}[!t]
\includegraphics[width=8cm]{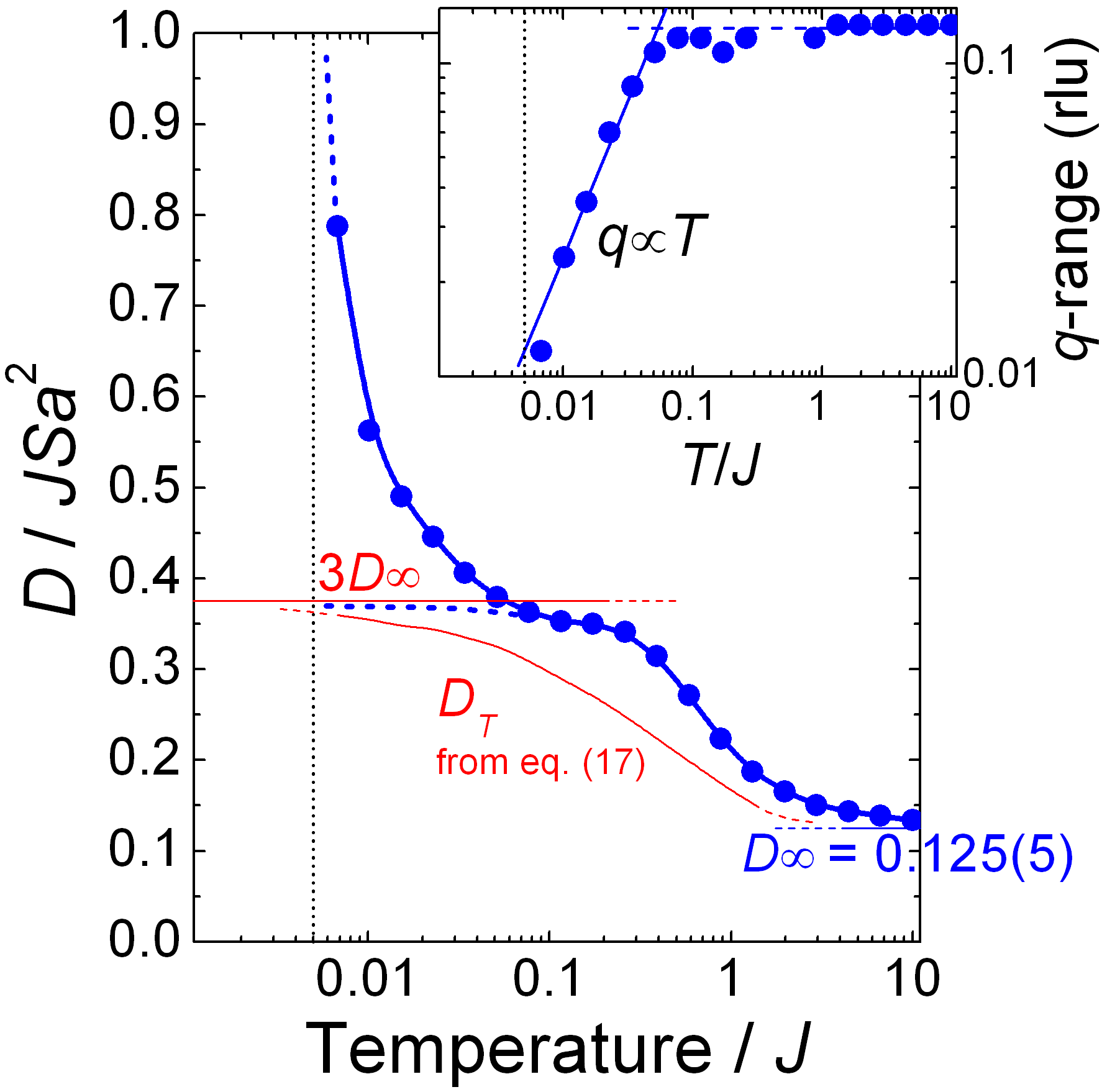}
\caption{(Color online) Diffusion coefficient and wave-vector range of
  validity of the diffusive approximation (inset) as a function of
  temperature. The red lines are the predictions obtained by the
  different models (see text); the value of $D_\infty$ is obtained by
  extrapolating the numerical data from $T/J=1$ to $10$; and the
  vertical dotted black line at $T/J=0.005$ is for the transition
  toward coplanarity.}
\label{fig:diffusion2}
\end{figure}

We proceed as follows to extract $D_T$. The relaxation time
$\tau_T(q)$ is obtained by fitting the scattering function
$S(\mathbf{Q},t)$ at some given temperature and wave-vectors using
Eq.~(\ref{eq:a:4}) for times $t\gg t_{vc}$: an example of such a fit
is presented on Fig. \ref{fig:diffusion1} (a) for $T/J=0.17$ at
wave-vectors taken along the $[h,h]$ direction around $q=0$ (in this
model the diffusive behavior is isotropic in $\mathbf{Q}$-space, so
other directions are not represented in the figure). Then, fitting the
relaxation time $\tau_T(q)$ versus $q^2$ for each temperature gives
both the range of validity in $q$ of the diffusive behavior and the
diffusion coefficient $D_T$ (see Fig. \ref{fig:diffusion1} (b)).

The temperature dependence of $D_T$ is represented in
Fig.~\ref{fig:diffusion2} in both paramagnetic and cooperative
regimes. At high temperature, $D_T$ asymptotically tends to a constant
value $D_\infty = 0.125(5)\ JSa^2$. This value is a little higher
than the prediction\cite{Marshall1968}
\begin{eqnarray}
D_T & = &  \frac{\pi}{2\sqrt{3}}
\lim_{q\rightarrow 0}
\left( \frac{\langle \omega^2 \rangle}{q^2}\right)
\left( \frac{\langle \omega^2 \rangle}{\langle\omega^4\rangle}\right)^{1/2}
\label{eq:diffus_LTE}
\end{eqnarray}
which is obtained by considering a Lorentzian response for
$S(\mathbf{Q},\omega)$ truncated at frequencies $\omega t_{vc}> 1$ to
take into account the failure of the exponential approximation at
times shorter than $t_{vc}$ (note that the coefficient
$\frac{\pi}{2\sqrt{3}}$ becomes $\sqrt{\pi/2}$ if we consider a short
time expansion instead of a rough cut-off, although the global
expression remains identical \cite{Lurie1974,Huber2003}). In the
infinite temperature limit, expression (\ref{eq:diffus_LTE}) gives
$D_\infty \simeq 3/16 r_0^2 (J/\hbar) \sqrt{zS(S+1)}$, with $r_0 =
a/2$ the distance between two nearest neighbors and $z=4$ the
connectivity of the kagome lattice. Using $JS(S+1)/\hbar\rightarrow
JS^2$ for classical spins, we find that $D_\infty = 3/32 \ JSa^2
\simeq 0.09375\ JSa^2$. This small discrepancy between numerical and
theoretical results was already noticed in 1D systems, and
is associated with the failure of the short time expansion which should be
carried to higher orders\cite{Lurie1974}.

For $T\lesssim J$, it becomes more difficult to obtain a simple theory
since other processes appear beside spin diffusion. However, by
considering the temperature dependence of the constant ratio $(\langle
\omega^2 \rangle / \langle\omega^4 \rangle)^{1/2}$ in the whole
temperature range, it is possible to rewrite Eq.(\ref{eq:diffus_LTE})
as a function of the macroscopic susceptibility and internal
energy\cite{Huber2003}
\begin{eqnarray}
D_T & \propto & U(T)/\chi(T).
\label{eq:DT}
\end{eqnarray}
Although the $O(N)$ model does not reproduce quantitatively the
simulations, it is possible to capture the global shape of the
diffusion coefficient above $T/J=0.05$ (see red curve in
Fig. \ref{fig:diffusion2}) using the analytic expressions of $U(T)$
and $\chi(T)$ derived in ref.[\onlinecite{Garanin1999}]

Finally, in the very low temperature limit $T\ll J$, the
infinite-component spin model coupled to a Langevin dynamics (see
Sec.~\ref{sec:diffusion} and Appendix \ref{appendix:ncomp}) predicts a
temperature independent diffusion
coefficient. Fig.~\ref{fig:diffusion2} shows that $D_T$ reaches a
plateau below $T/J=0.1$ at around $0.37(1)\ JSa^2$. Moreover, from
ref.~[\onlinecite{Garanin1999}] and using expression (\ref{eq:DT}), we
obtain a ratio $(D_{T=0}/D_\infty)_{O(N)}=3$ between zero and
infinite temperature. This quantitatively agrees with the value
$2.8(4)$ determined from our simulations, while extrapolating the
value of the plateau down to $T=0$ from the behavior observed around
$T/J\sim 0.1$ (see figure \ref{fig:diffusion2}).
At lower temperature $T/J<0.05$, the $O(N)$ model rapidly fails since
it does not capture the entropic selection of the coplanar states.  The
diffusion coefficient seems to diverge when the temperature reaches
the octupolar transition, while the wave-vector range of validity of
 spin diffusion, which is restricted at low temperatures by the
condition $q \xi < 1$ with $\xi$ the correlation length
\cite{Lurie1974}, shrinks to very small wave-vectors (see inset of
Fig. \ref{fig:diffusion2}).

From our simulations, it is not possible to state that the
long-wavelength diffusive behavior disappears in the coplanar regime
in favor of propagative spin transfers or if it is simply reduced to a
$q$-range smaller than the resolution $\delta q = (Na)^{-1}$, denoting
that the correlation length becomes larger than the lattice size. In
this latter case, bigger lattices should be considered to avoid
finite-size effects.
In any case, diffusive behavior may exist even in a long-range ordered
AFM as long as non-linear effects such as interacting spin waves
are significant. These interactions are particularly strong in
frustrated magnets even at very low
temperatures\cite{Starykh2006,Mourigal2013}, so spin diffusion may
still be present in the coplanar regime, although being limited to
very long wavelengths and negligible in intensity compared to
propagative spin transfers.
\begin{figure*}[ht]
\hfill{}
\includegraphics[width=14cm]{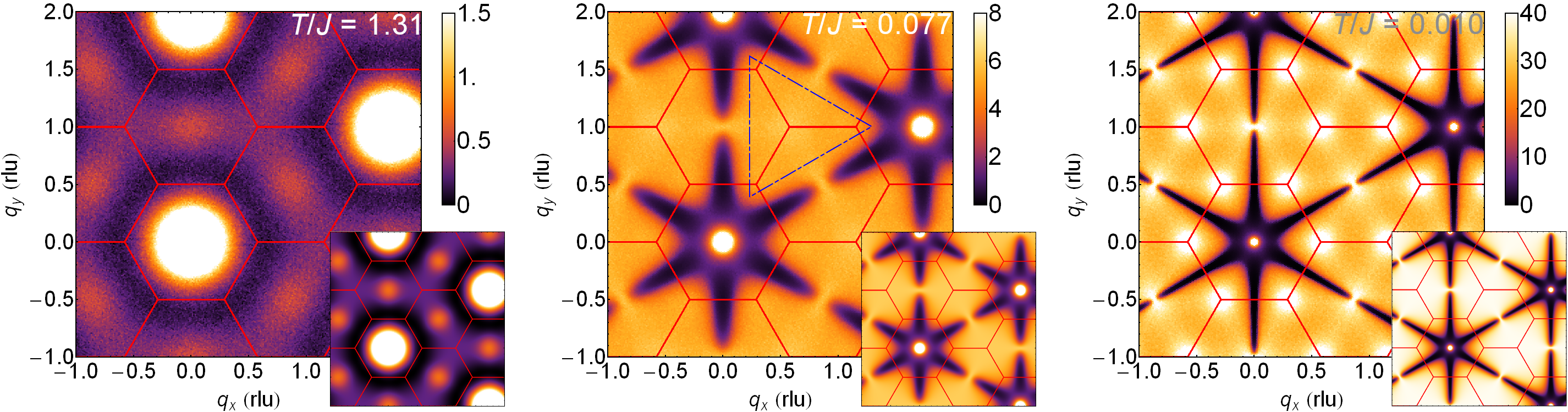}
\hfill{}
\caption{(Color online) Temperature dependence of the relaxation time
  $\tau({\bf Q},T)$ in the cooperative regime. Intensity maps of the
  relaxation time $\tau(\mathbf{Q},T)$ in reciprocal space for
  $T/J=1.31$ (a), $0.077$ (b) and $0.010$ (c), extracted by fitting
  the scattering function $S(\mathbf{Q},t)$ at each wave-vector using
  equation (\ref{eq:relaxmax}). The prediction of the
  infinite-component spins model stands at the bottom right for each
  temperature. The red lines bound the Brillouin Zones.}
\label{fig:relax1}
\end{figure*}

\subsubsection{Relaxation at generic wave-vector : lifetime of the ground states}
\label{sssec:relaxmax}

\begin{figure}[!b]
\centering
\includegraphics[width=8cm]{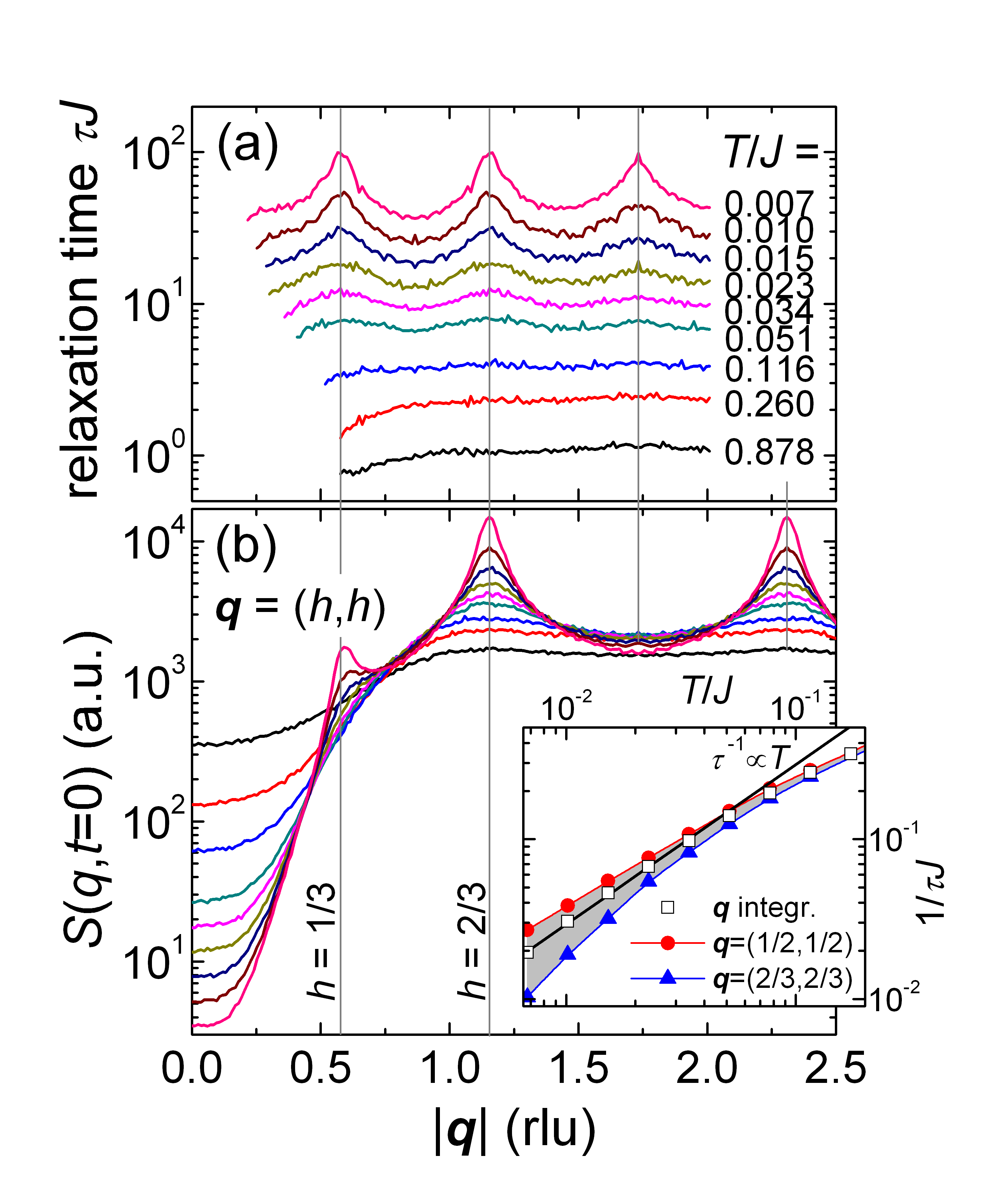}
\caption{(Color online) Influence of the temperature on $S(q,t)$. (a)
  Relaxation time and (b) instantaneous scattering function
  $S(\mathbf{Q},t=0)$ versus $\mathbf{Q}=(h,h)$ at many temperatures
  in the paramagnetic and cooperative regimes. The vertical gray lines
  at $h=n/3$ are for BZ vertices ($n=1,2,4$) and center
  ($n=3$). (inset) Evolution of the relaxation time with temperature
  at some high symmetry positions in reciprocal space. The open squares
  represent the relaxation time obtained after integration over
  wave-vectors away from nodal lines where $\tau$ is predicted to be
  $q$-independent from ICSM (see text), with a $1/T$ law which is
  represented in black.}
\label{fig:relax2}
\end{figure}

The autocorrelation function $A(t)$ gives useful information about the
global relaxation of the system and may be an efficient way to probe
the evolution of the stiffness with the development of correlations at
low temperature.  In a previous study \cite{Robert2008} as well as in
fig~\ref{fig:autocorr} of this paper, it is shown that a decaying
exponential qualitatively describes the autocorrelation function in
the paramagnetic and cooperative paramagnetic regimes $-$ at least in
a certain time range, this range being highly reduced in the
paramagnetic regime because of the $1/t$ diffusive tail.
Thus, the wave-vector-averaged relaxation time can be extracted for
each temperature, its inverse $\Gamma_r(T) = \tau_r(T)^{-1}$ being
represented in Fig. \ref{fig:tempsrelax}. The relaxation rate
$\Gamma_r(T)$ goes from the constant value $J$ in the paramagnetic
regime, to an algebraic law $\mathcal{A} T^{\alpha}$ in the
cooperative regime with $\alpha=0.94(3)$ close to one, reflecting
a slowing down of the spin fluctuations.

For comparison, note also that a similar result has been obtained in
the pyrochlore antiferromagnet using simulations and phenomenological
arguments\cite{Moessner1998,Conlon2009}.

Nevertheless, $A(t)=\int{\mathrm{d}^2\mathbf{Q} S(\mathbf{Q},t)}$ only
provides qualitative information in the case of wave-vector dependent
fluctuations. So it is necessary to study the $\mathbf{Q}$-resolved
scattering function $S(\mathbf{Q},t)$ to understand the overall
dynamical properties of the system.  In the following, we assume an
exponential decay of the scattering function at each $\mathbf{Q}$
\begin{eqnarray}
S(\mathbf{Q},t) & = & S(\mathbf{Q},0)\mathrm{e}^{-t/\tau_T(\mathbf{Q})}
\label{eq:relaxmax}
\end{eqnarray}
and use the same treatment as the one discussed in the previous
section for long wavelengths. The relation (\ref{eq:relaxmax}) was
checked to be a good approximation at most wave-vectors. In particular
Eq.~(\ref{eq:relaxmax}) is justified in the paramagnetic and
cooperative regimes because the long time dynamics is dominated by
relaxation processes away from nodal lines. In that case,
propagating excitation can be neglected in the first
approximation\cite{Robert2008}. So, extracting the relaxation time
$\tau_T(\mathbf{Q})$ from the numerical data allows us to distinguish
the dynamical properties of short-range correlated domains having a
propagation-vector $\mathbf{Q}$ at temperature $T$.

Fig.\ref{fig:relax1} displays maps of $\tau_T(\mathbf{Q})$ in
reciprocal space for temperatures $T/J=1.31$, $0.077$ and $0.01$; the
red lines represent the BZ edges. For comparison, the same maps
obtained from the $O(N)$ model (Sec.~\ref{sec:diffusion} and Appendix
\ref{appendix:ncomp}) are shown in insets. Cuts of these maps as well
as the instantaneous scattering function $S(\mathbf{Q},t=0)$ are shown
in Fig. \ref{fig:relax2} along the $[h,h]$ direction for several
temperatures.

In the cooperative regime (Fig. \ref{fig:relax1} (b)), the relaxation
time seems to be nearly independent of wave-vector in regions
away from the nodal lines and the zone center where the dynamical
properties are dominated by diffusion and spinwave like
processes. This result is very similar to the predictions of the
$O(N)$ model and is rather intuitive in the light of
Ref.\cite{Halperin1977} since no particular correlations are favored
in this temperature range, all locally ordered domains having roughly
the same relaxation time. More generally, it is striking that the
$O(N)$ model reproduces accurately the overall $q$-dependence of the
relaxation time above $T/J=0.05$.

The simulations show that $\tau_T(\mathbf{Q})$ (Fig. \ref{fig:relax1}
(c)) becomes more structured at lower temperatures.  In particular, it
is clear from Fig. \ref{fig:relax2} (a) and (b) that the longest
relaxations coincide with the correlations peak of the
\s3s3 phase that are located around BZ vertices. This slowing down of
the spin fluctuations at BZ vertices, obviously not reproduced by the
$O(N)$ model which does not take into account order-by-disorder
phenomena, is thus observed at higher temperatures than the onset of
\s3s3 static correlations, which occurs only when
$T/J\lesssim0.005$\cite{Zhitomirsky2008}.

The temperature evolution of the relaxation time close to and away
from the BZ vertices is represented in blue and red in the inset of
figure \ref{fig:relax2} (b). The shaded region between these two
curves symbolizes the relaxation time distribution (for wave-vectors
contained in the blue triangle of Fig. \ref{fig:relax1} (b)). The open
black squares are the mean value of this distribution. Although the
$O(N)$ model neglects the wave-vector dependence of
$\tau_T(\mathbf{Q})$ below $T/J=0.05$, the wave-vector-averaged
relaxation time is roughly consistent with the law
$\tau_T(\mathbf{Q})\propto T^{-1}$ obtained in sec.\ref{sec:diffusion}
at low temperatures.

To conclude, these results suggest that the fluctuations around the
ground state manifold are strongly affected by the entropic selection
far above the transition toward coplanarity which occurs only at
$T/J=0.005$.
Contrary to the antiferromagnetic Heisenberg model on the pyrochlore
lattice\cite{Conlon2009}, the $O(N)$ spins model describes only
qualitatively both the diffusive and higher wave-vectors regimes in a
restricted temperature range $0.05<T/J\ll 1$.
%

\section{coplanar regime}
\label{ssec:coplanar}

\begin{figure}[!t]
\centering
\includegraphics[width=8cm]{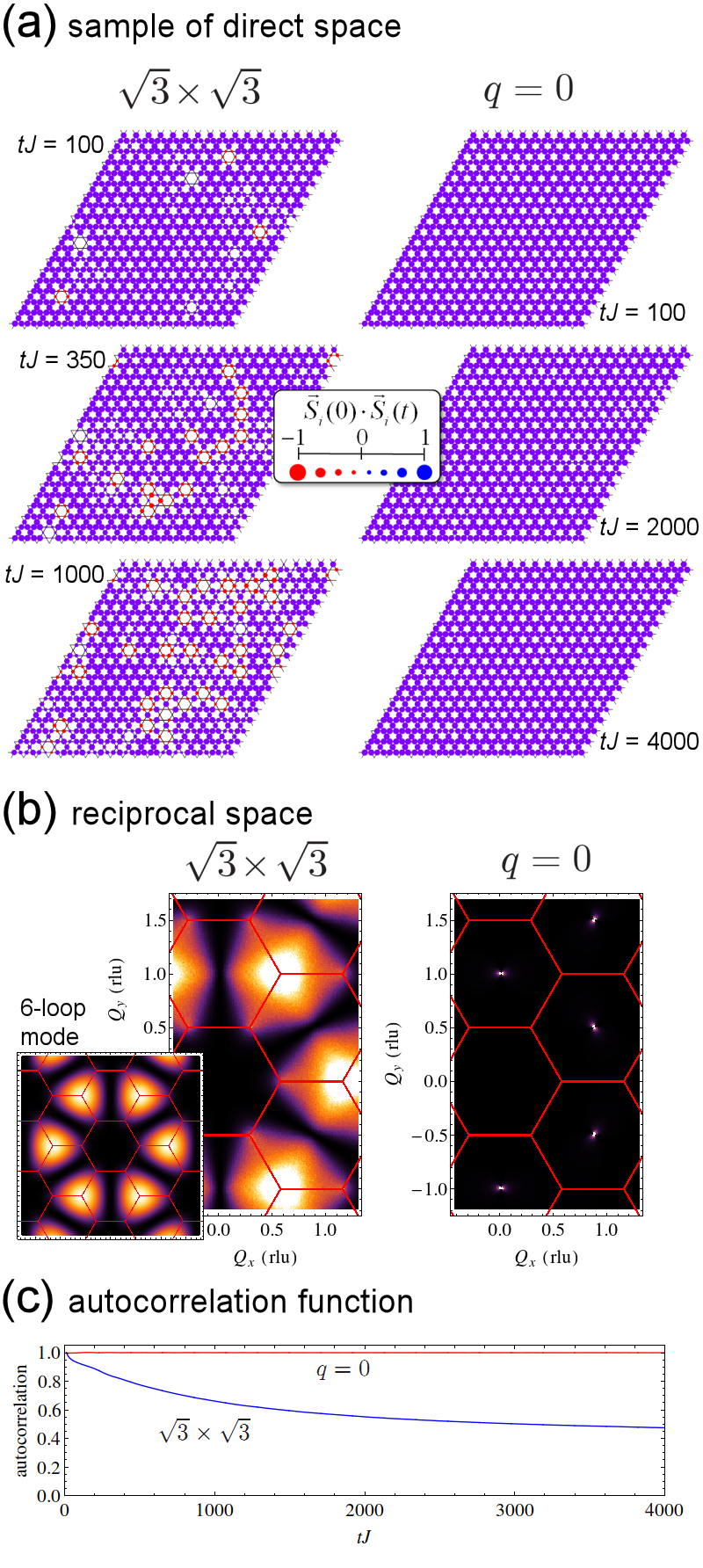}
\caption{(Color online) Dynamics of the weatherwane defects for the
  uniformly randomized spin configurations in the $q=0$ and \s3s3
  states seen from the correlation functions \mbox{${\bf s}_i(t)\cdot {\bf
    s}_i(0)$} ($\Delta E\simeq 6.4\,10^{-4}\ J$). (a) autocorrelation in
  direct space at times $tJ=100$, 350 and 1000 (resp. $tJ=100$, 2000
  and 4000) for the \s3s3 (resp. $q=0$) spin configuration. Blue
  (resp. red) disks are for positive (resp. negative) autocorrelation
  $\mathbf{s}_i(0) \cdot \mathbf{s}_i(t)$ at site $i$. (b) Resulting
  static scattering function in reciprocal space of both spin
  configurations after numerical integration. For comparison, the
  bottom left inset shows the structure factor of one weathervane
  mode. (c) Time evolution of the global autocorrelation function.}
\label{fig:softmodes}
\end{figure}

\subsection{Low temperature landscape}
\label{ssec:gen-biblio-predictions}

The goal of this section is to motivate and justify the investigation
of the $\mathbf{Q}$-resolved dynamical scattering function in the time
range of interest.

In the coplanar (octupolar) regime\cite{Chalker1992,Zhitomirsky2008},
the incoherent spin dynamics induced by thermal fluctuations is not
the only channel of relaxation.
In this low temperature phase, the low energy manifold can be thought
as a neighborhood of all 3-colorings of the kagome lattice, which
form a discrete manifold, all colorings being globally related one to
another by continuous bridges, the collective rotation of spins
belonging to 2-colored loops of the discrete manifold.
As a consequence, this low energy manifold is connected, even though
it must be noted that from the discrete point of view, only the
dynamics within the discrete 3-colors manifold induced by the 2-color
loop move leads to a non-connected structure, the manifold being
split into distinct Kempe sectors\cite{Mohar2010,Cepas2012}, {\it i.e.} spin
configurations which cannot be related one to another through loop
moves only.
In other words, it is only possible to go from one sector to another
through the use of a defect, a highly unlikely event at low
temperature.
Consequently, in a typical time scale of $tJ < 1000$, one can consider
that the system is trapped in a Kempe sector and does not escape it.

Whatever the sector the system is trapped in, there exists loops of
different lengths $p=2+4n$ with $n>0$\cite{Henley1993,Ritchey1993}.
Using periodic boundary conditions, loops can be divided into two
categories; winding and non winding loops.
One may expect different dynamics for these two families.
Actually, we will now see that at low temperature, the microscopic
spin model we are interested in, in the time range of interest,
discriminate even more drastically within each family.

Let us consider two archetypal 3-coloring, the well known long range
ordered $q=0$ and \s3s3 spin configurations whose shortest weathervane
modes are respectively infinite lines and small loops of 6 spins.
In order to mimic a very low temperature regime, we introduce a small
amount of energy in each phase, by uniformly randomizing each spin
configuration with a $\Delta E\simeq 6.4\,10^{-4}\ J$
.
Then, equations of motions are integrated and time evolution for each
case is represented in Fig. \ref{fig:softmodes}, with (a), the
autocorrelation of each spin in direct space and (b), the associated
static structure factor $S(\mathbf{q},\omega=0)$.
While hexagonal loops are activated and their number increases with
time for the \s3s3 spin configuration, no flipped loop is detected
for the $q=0$ phase.
Note also that for the \s3s3 spin configuration, {\it no loop} of
length greater than 6 occurs.
loop of length $L>6$ as well as infinite loops are therefore absent
{\it at this time scale} and do not play any role in the dynamical
properties.
In reciprocal space, this results in a negligible diffuse spectral
weight at $\omega = 0$ for the $q=0$ phase in opposition to the \s3s3
phase (see Fig. \ref{fig:softmodes} (down)).

Because it is now well established that entropy stabilizes \s3s3
correlations at low temperatures, one may expect that
thermodynamically, spin configurations belong to the corresponding
Kempe sector.
On the time scale of the simulations, one may consider the phase space
to be the 6-loops neighborhood of this configuration, {\it i.e} all
accessible configurations starting from the pure \s3s3 phase and
applying non-overlapping 6-loop move, keeping in mind that all
operations are not commutative.

Therefore, while in reciprocal space \s3s3 spin pair correlations
gives rise to sharp peaks located at the Brillouin zone vertices,
their width being inversely proportional to the correlation length
$\xi_{\sqrt{3}}(T)$, the presence of flipped hexagonal loops yields 
an elastic diffuse spectral weight in reciprocal space, since the
presence of those 'defects' in the parent periodic structure requires
an infinite number of Fourier components.
The form factor of one such 'defect' is represented in inset of
Fig. \ref{fig:softmodes} (b) and indeed results in broad bumps softly
stretched along BZ edges.
Consequently, the instantaneous structure factor is expected to be a
superposition of both sharp and broad features located at different
regions of reciprocal space, and whose origin are of different nature
: probing the dynamics at different wave vectors will give
information on different relaxation processes.

\subsection{Models, predictions about time scales}
\label{sssec:predictioncop}

The analytic approach described in Sec.~\ref{sec:diffusion}
obviously fails to describe the fluctuations around such a ground
state manifold, simply because it neglects the order by disorder
phenomena occurring at very low temperature. However, using qualitative
arguments, it is possible to roughly predict how the dynamical
properties would evolve in the presence of an entropically induced
potential well.

In a first approximation, let us consider the time evolution of a
single loop diffusing in such a landscape, whose dynamics is described
by a simple stochastic differential equation. In this approach, we
also ignore the interactions between the weathervane modes and the
spin waves sensitive not only to the ground state manifold (in the
sense of internal energy), but also to the excitation spectrum.

This dynamics should have, at sufficiently low temperatures ($T\ll
V_L$ with $V_L$ the height of the free energy barrier), two
distinct timescales, corresponding to $(i)$ the required time to
overcome the barrier and flip the loop, $(ii)$ the weak loop
fluctuations around the plane of coplanarity. While the latter time
scale will mainly affect the out-of-plane component for sufficiently
small fluctuations ({\it i.e.} at sufficiently low temperatures), flipping a
loop will also influence the in-plane channel since such a motion
induces a change of three-coloring.

Classically, the in-plane relaxation associated with loop motions should
obviously undergo a reduction of the number of flips with decreasing
temperature, described by the activation law
\begin{eqnarray}
\tau_\parallel & = & \tau_0\exp(-V_L/T).
\label{eq:actlaw}
\end{eqnarray}
An estimation of the energy barrier height $V_L$ has been obtained
within Gaussian spin-wave
theory\cite{Cepas2011,Ritchey1993,Chandra1993}. In particular, it was
shown that the $\pi$-periodic potential well induced by quantum
fluctuations has the form $V(\phi)=\eta L |\sin(\phi)|$, with
$\eta=0.14$\cite{Ritchey1993,Henley1993,Cepas2011} and $\phi$ the
angle between the ``averaged'' coplanar spin plane and the plane
defined by the spins of the considered loop. Therefore, in the low
temperature limit where quantum fluctuations dominate,
$V_L=V(\phi=\pi/2)\propto L$ only depends on the loop length
$L$. However, in the presence of substantial thermal fluctuations
(classical limit), the barrier height is renormalized
$V_L=TL\log(2\eta J S /T)$\cite{Ritchey1993,Chandra1993}. Combining
this latter expression with equation (\ref{eq:actlaw}) leads to the
power law
\begin{eqnarray}
\tau_\parallel & = & \tau_0 \left(\frac{2\eta J S}{T}\right)^{L},
\label{eq:algdep}
\end{eqnarray}
whose exponent is the loop length.

The behavior of the second time scale, {\it i.e.} weak fluctuations of the
loops within the entropic potential well, strongly depends on the
precise shape of the well and is more difficult to handle. Indeed, for
small angle, the fluctuations of the neighboring loops cannot be
neglected anymore. In particular, it was shown that taking into
account these loops, interactions round out the well bottom such that
its $\phi$ dependence becomes quadratic for $\phi\lesssim\phi_0$, with
$\phi_0 = \langle\phi^2\rangle^{1/2} = \sqrt{T/J}$ the rms induced by
thermal fluctuations\cite{Henley1993}.
The stochastic Langevin equation in a quadratic well
$V(\phi)=d(T)\phi^ 2$, with $d(T)$ an effective onsite planar
anisotropy coefficient which possibly depends on the temperature, may
be solved analytically. This leads to a relaxation time
$\tau_\perp\propto 1/d(T)$, pointing out that the temperature dependence of
the out-of-plane relaxation time follows that of $d(T)$ : for
instance, a temperature independent well bottom would lead to a
constant out-of-plane relaxation time, leading to no freezing effects
down to $T=0$ (in this particular case, this would actually be the
amplitude of the fluctuations which would tend to zero with decreasing
temperature).

In the next section, we numerically test these ideas and try in
particular to prove the presence of several characteristic time scales
in the spin dynamics. We also qualitatively discuss the possible role
of the interactions between weathervane and spin waves modes, which
break the 120$^\circ$ rule between neighboring spins and lead to
incoherent spin fluctuations.

\subsection{Numerical results and discussion}
\label{ssec:numcop}

\begin{figure}[t]
\centering
\includegraphics[width=8cm]{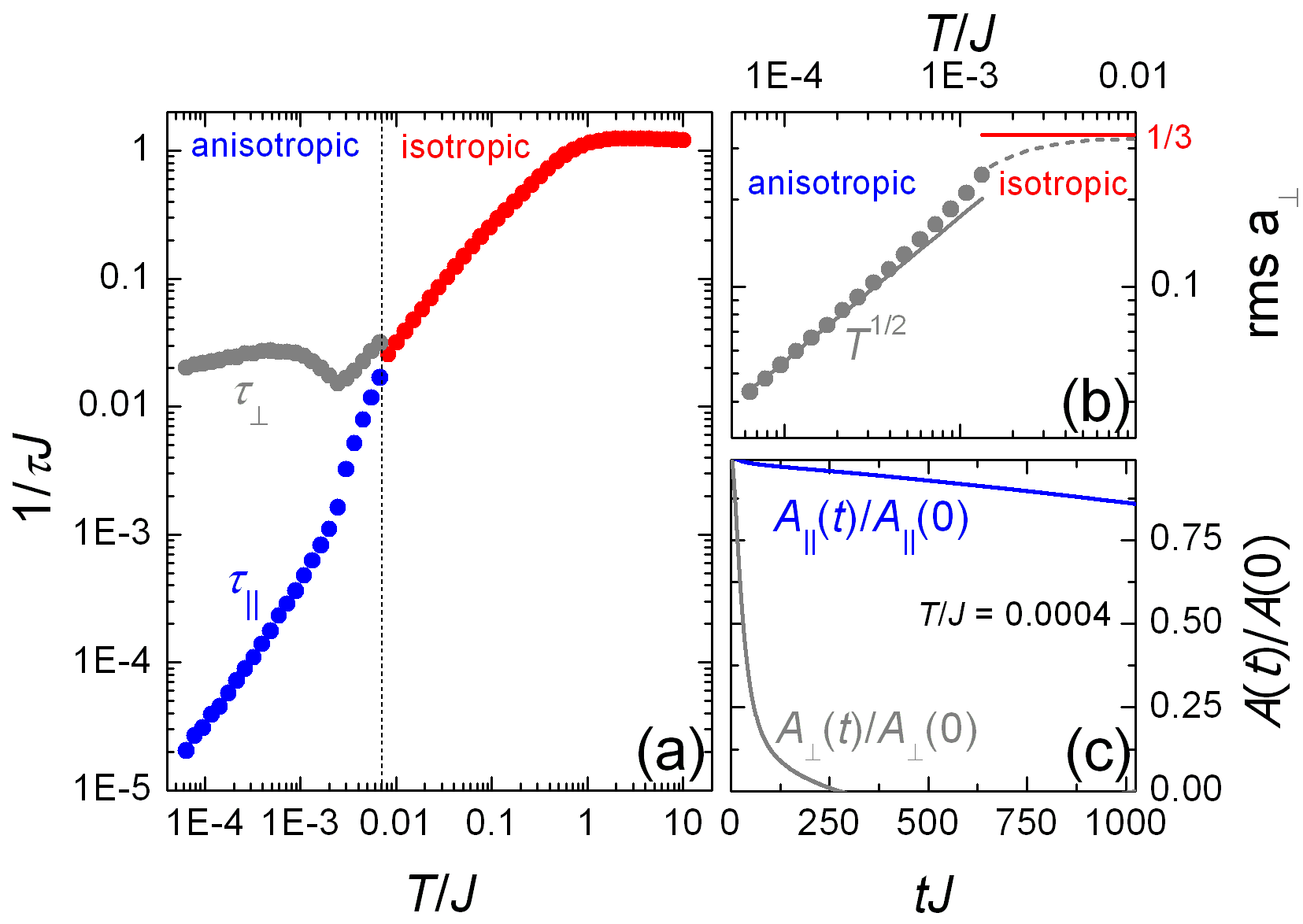}
\caption{(Color online) Algebraic temperature dependence of Relaxation
  time in the three different regimes. (a) Relaxation rate $1/\tau J$
  {\it vs.} temperature: red is for the isotropic high temperature
  regimes (paramagnetic and cooperative) while blue and gray
  respectively corresponds to the in-plane and out-of-plane
  fluctuation rates. (b) Out-of-plane amplitude of spin fluctuations
  in function of temperature. (c) In-plane/out-of-plane separation of
  the autocorrelation function for $T/J=0.0004$.}
\label{fig:tempsrelax}
\end{figure}

The autocorrelation function $A(t)$ plotted on figure
\ref{fig:autocorr} exhibits different behaviors depending on the
considered temperature range : paramagnetic, spin liquid or coplanar
regimes.
Although the autocorrelation in the spin liquid phase can be described
by a single decaying exponential, such treatment is not valid (see
Sec.\ref{ssec:paraliqu}) far above the transition toward coplanarity
(for $T/J\lesssim 0.05$)$-$. Below the crossover temperature,
Fig.\ref{fig:autocorr} (c) shows a more complex behavior with at least
two time scales separated by a crossover at around $tJ\simeq60$.
Since the coplanar regime is (by definition) anisotropic in spin
space, different dynamical behaviors are expected for the in-plane and
out-of-plane components, each one being likely associated with
different relaxation processes.
Separating the two contributions $A_\parallel(t)$ and $A_\perp(t)$ of
the autocorrelation function in our simulations appears natural. The
short time scale ($tJ<60$) is governed by out-of-plane relaxation,
while the in-plane relaxation governs the long time regime (see figure
\ref{fig:tempsrelax} (c) for $T/J=0.0004$).
From these considerations, $A(t)$ can be split into two exponential
contributions below $T/J\lesssim 0.005$
\begin{eqnarray}
A(t) & = & a_\perp \mathrm{e}^{-t/\tau_\perp} + a_\parallel \mathrm{e}^{-t/\tau_\parallel},
\label{eq:Atparaperp}
\end{eqnarray}
with $\tau_{\perp}\ll\tau_{\parallel}$ the relaxation times and
$a_{\perp / \parallel} = \frac{1}{N} \sum_i (\mathbf{S}_i^{\perp
  / \parallel})^2 \in [0,1]$ the amplitudes of the
out-of-plane/in-plane fluctuations such that $a_\parallel + a_\perp =
1$, gives a good agreement with the numerical data.

$\tau_\parallel$ behaves like the relaxation time obtained in the spin
liquid regime: it follows an algebraic law $\mathcal{A} T^{\alpha}$
with a slightly higher exponent ($\alpha \simeq 1.2(1)$) denoting a
slowing down of the spin fluctuations below the transition, the
coplanarity inducing a stiffness in the spin texture. This exponent
value remains qualitative and may slightly be influenced by finite
size effect. Meanwhile, $\tau_\perp \simeq 35(5)\, J^{-1}$ seems to be
roughly temperature independent.
These different dynamical behaviors come with a decrease of the
out-of-plane spin component $a_\perp \propto T^{1/2}$ (see
Fig. \ref{fig:tempsrelax} (b)), in agreement with the equipartition
theorem in the presence of out-of-plane quartic modes.

\begin{figure}[t]
\centering
\includegraphics[width=8cm]{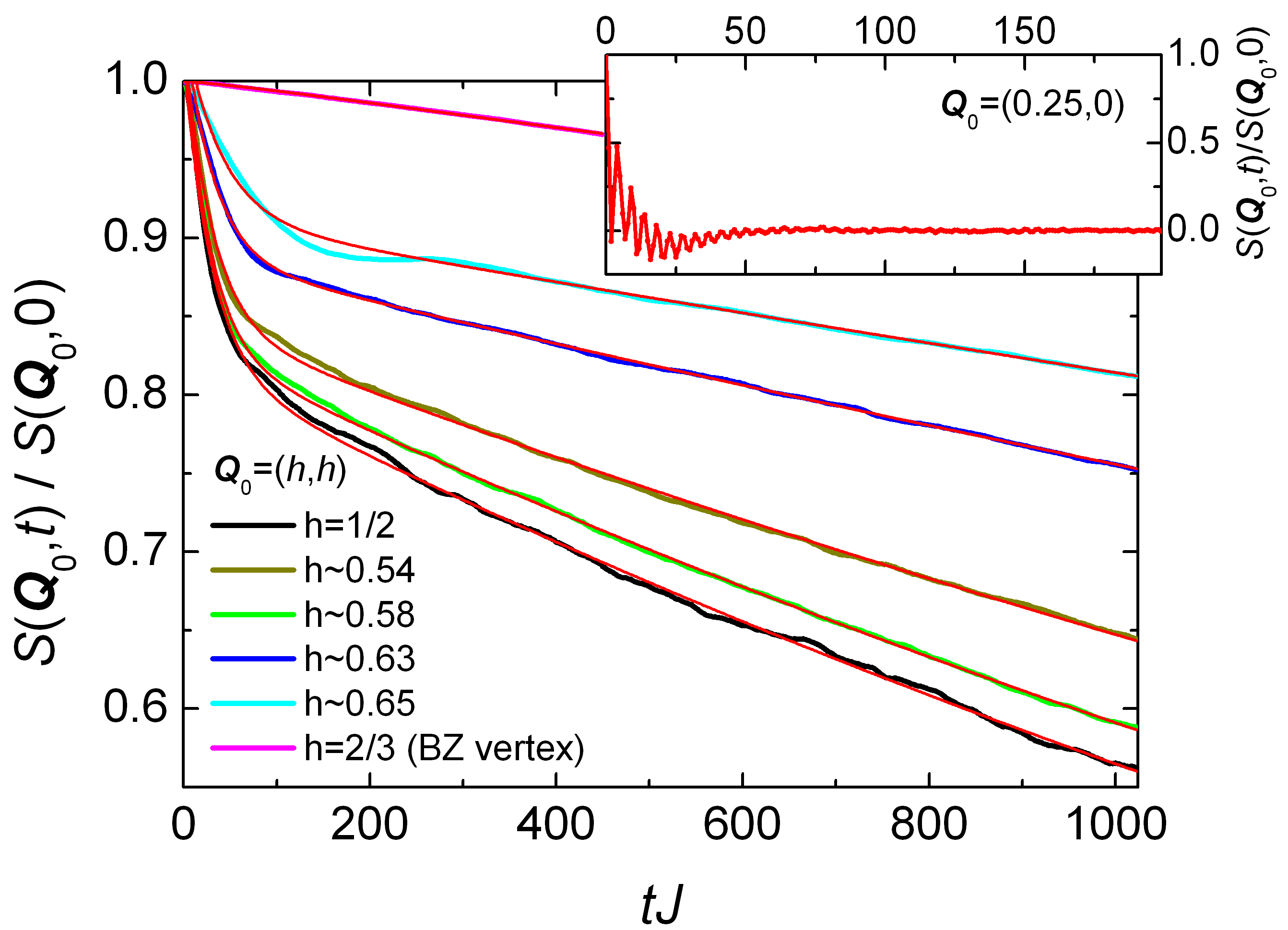}
\caption{(Color online ) Scattering function $S(\mathbf{Q},t)$ for different
  wavevectors at $T/J=0.0006$ as a function of time $tJ$. The fits
  obtained using equation (\ref{eq:SqtParaPerp}) at each wave vector
  are represented in red. Inset : Scattering function for
  $\mathbf{Q}=(0.25,0)$ along the nodal line pointing out finite
  frequency features.}
\label{fig:fitscoplanaires}
\end{figure}

\begin{figure*}[!t]
\hfill{}
\includegraphics[width=17cm]{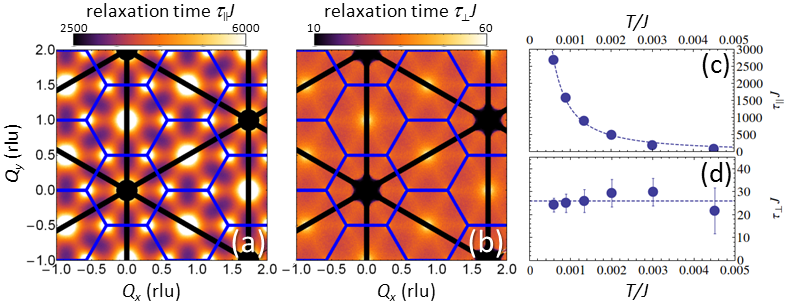}
\hfill{}
\caption{(Color online) $q$ dependence of the parallel and
  perpendicular relaxation times in the octupolar regime. Intensity
  maps of the parallel (a) and perpendicular (b) relaxation times
  $\tau_q^{\perp / \parallel}(\mathbf{Q},T)$ in reciprocal space for
  $T/J=0.0006$, extracted by fitting the scattering function
  $S(\mathbf{Q},t)$ at each wave-vector using equation
  (\ref{eq:SqtParaPerp}). The blue lines bound the Brillouin
  Zones. The black thick lines and disks hide the regions where the
  quasistatic scattering function vanishes (nodal lines), so that
  using Eq. (\ref{eq:SqtParaPerp}) is meaningless around these
  positions. (c) and (d), obtained around $\mathbf{Q}=(1/2,1/2)$,
  respectively show the in-plane and out-of-plane characteristic
  fluctuation time of the loops.}
\label{fig:relaxcoplanaire}
\end{figure*}

Considering the dynamical scattering function $S(\mathbf{Q},t)$, which
gives access to the wave vector dependence of the relaxation times,
yields more insight about the underlying mechanism leading to the
strongly different dynamical behaviors of the in-plane and
out-of-plane spin components : by avoiding the wavevector-averaging
effects, we can detect the regions of reciprocal space leading to such
a behavior.
As for the higher temperature results, $S(\mathbf{Q},t)$ is dominated
by quasistatic relaxation for wave vectors away from the nodal
lines\footnote{Note that this approximation is actually fully
  compatible with previous results pointing out the existence of
  propagative spin waves, as long as the intensity of the modes is
  small compared the static intensity $S(\mathbf{Q},\omega=0)$. The
  spin wave excitations contribute to the scattering function
  $S(\mathbf{Q},t)$ by a negligible modulation of high frequency (of
  the order of $J$).}. So following the in-plane/out-of-plane
separation performed for the autocorrelation, the scattering function
can be approximated for each $\mathbf{Q}$-value by
\begin{eqnarray}
S(\mathbf{Q},t) & = & S^\perp(\mathbf{Q},0) \mathrm{e}^{-t/\tau^\perp_q} + S^\parallel(\mathbf{Q},0) \mathrm{e}^{-t/\tau^\parallel_q}.
\label{eq:SqtParaPerp}
\end{eqnarray}
Some of those fits along the $Q=(h,h)$ directions are shown in
Fig. \ref{fig:fitscoplanaires}, first pointing out that the peculiar
shape of the autocorrelation function in the coplanar regime is not
induced by the wave-vector averaging, since the two characteristic time
scales are also observed at each wave-vector value.
Then, the relaxation times obtained by fitting the scattering function
at each wave vector using expression (\ref{eq:SqtParaPerp}) are
plotted as intensity maps in reciprocal space for $T/J=0.001$ on
Fig. \ref{fig:relaxcoplanaire} (a) for $\tau^{\perp}_q$ and (b) for
$\tau^{\parallel}_q$.

Handling the in-plane scattering function is a bit tricky, seeing that
the static spectral weight is a combination of two components with
very different signatures in reciprocal space as discussed in
subsection \ref{ssec:gen-biblio-predictions} : different relaxation processes
or lifetimes are probed depending on the wavevector value.
Around the BZ vertices, the static spectral weight is overwhelmed by
the sharp peaks resulting from \s3s3 correlated domains. Thus, the
time evolution of the scattering function around these positions
unveil the lifetime of these locally ordered states, which, from our
simulations, seems to diverge with decreasing temperature.
However, the static and dynamical properties around BZ vertices are
strongly affected by finite size effects at such low temperatures
since the correlation length $\xi_{\sqrt{3}}(T)$ may reach the lattice
size. Consequently, it is not possible in the current work to
quantitatively describe the temperature evolution of the relaxation of
\s3s3 correlations. 

On the other hand, the relaxation of the diffuse spectral weight at
generic wavevectors is representative of loop dynamics. The
corresponding time scales of those local motions in direct space are
almost independent of the system size for a sufficiently large number
of spins.

The in-plane components provide the average time to flip hexagonal
loops in given \s3s3 domains (which is different from the flipping
motion by itself which has already been discussed in a previous
article\cite{Robert2008}). Indeed, the in-plane spin correlations are
at the first order not sensitive to small fluctuations of the loops,
and full loop flips are naturally needed to alter three-coloring
states. Figure \ref{fig:relaxcoplanaire} (a) points out that the
averaged time to flip the loops is smaller than the lifetime
of the \s3s3 correlated domains. Each weathervane loop may be flipped
several times before the \s3s3 correlated domains to be relaxed.

However, since flipping an hexagonal loop requires to overcome the
free energy barrier separating the two neighboring three-colorings
(the ones before and after the flip), a depletion of the flipping
events with decreasing temperature is naturally expected. According to
Eq.~(\ref{eq:algdep}), the required time to flip a loop follows an
algebraic law $\tau(T)\propto T^\alpha$ with $\alpha$ is equal to the
loop length.
Numerical data obtained around $\mathbf{Q}=(1/2,1/2)$~rlu, shown in
figure \ref{fig:relaxcoplanaire} (c), are in very good agreement with
a power-law behavior
but the fitted exponent $\alpha\simeq 1.5(2)$ is around four time
smaller than the prediction $\alpha=6$ for hexagonal loops.
This discrepancy could be due to the interactions between the local
(loops) and non-local (spin waves) modes, which have been neglected
and probably lead to significant incoherent thermal fluctuations.
Note also that finite size effects although strongly reduced far away
from BZ centers and corners (see sec. \ref{sec:model}) can not be
totally excluded.

Away from the high symmetry directions of the Brillouin zone, the
characteristic time-scale of the out-of-plane fluctuations seems
roughly flat with wavevector. This result is consistent with the local
spin motions in direct space (see Fig. \ref{fig:relaxcoplanaire} (b))
and suggests that the relaxation is mediated by the loop fluctuations
at very low temperatures. $\tau^{\perp}_\mathbf{Q}$ weakly depends on
temperature and is around $28 J^{-1}$ for $T/J\lesssim0.005$ (see Fig.
\ref{fig:relaxcoplanaire} (d) at wave vector $Q=(1/2,1/2)$~rlu), as
previously noticed for the $\mathbf{Q}$-integrated scattering function
$A(t)$.

The presence of temperature independent spin fluctuations is
remarkable for a classical system, whose dynamics generally slows down
when the temperature decreases. It is however consistent with loops
slightly fluctuating around the plane of coplanarity, if considering a
temperature independent entropic well bottom $V(\phi)=d\phi^ 2$ (see
sec. \ref{sssec:predictioncop}). Nevertheless, to go beyond these
phenomenological considerations and confirm these numerical results,
theoretical predictions of the precise temperature dependence of the
entropic well are necessary in the limit of small angles $\phi$.

\begin{figure*}[!t]
\hfill{}
\includegraphics[width=17cm]{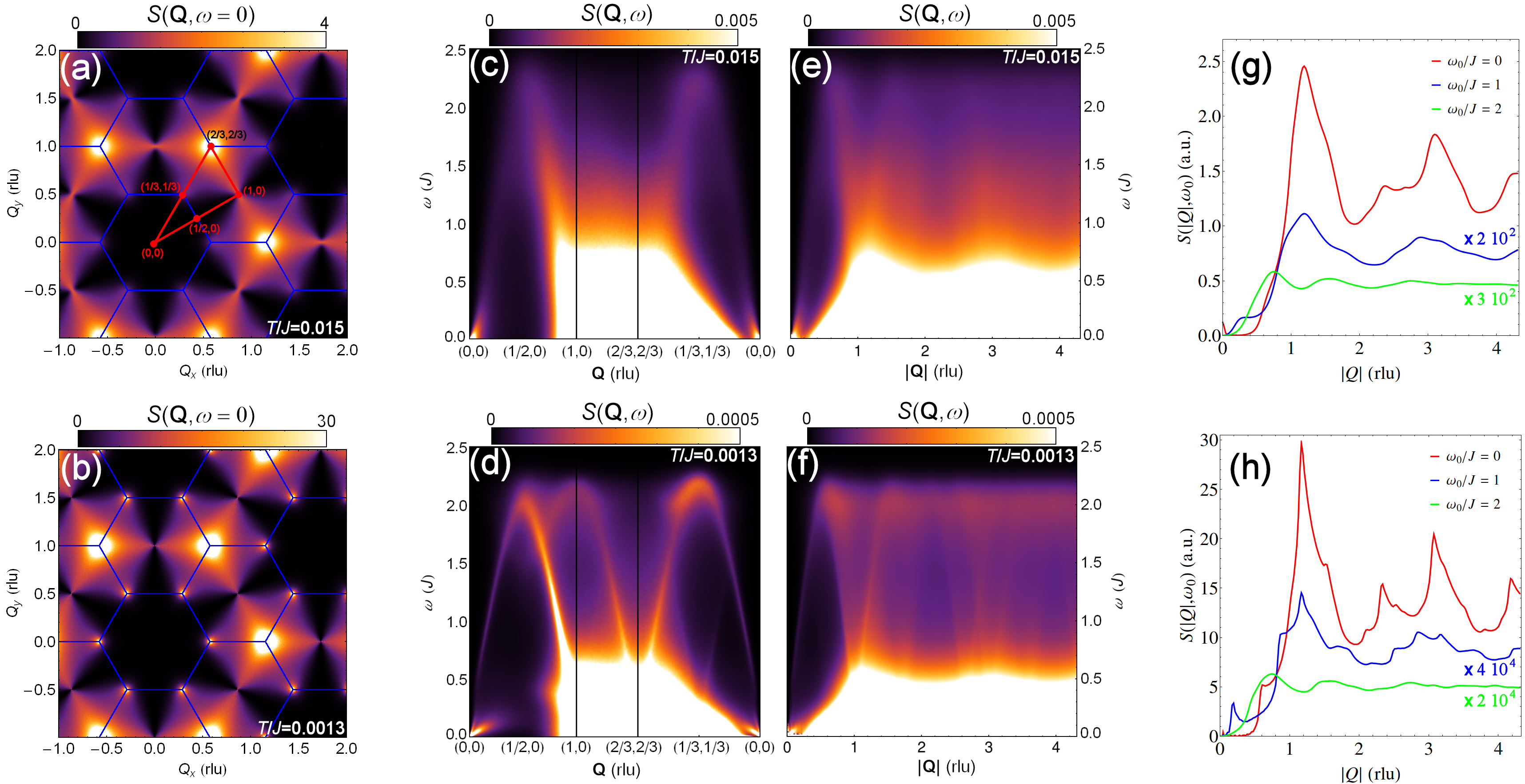}
\hfill{}
\caption{(Color online) Powder averaged scattering function in the
  octopular regime. (a,b) Static scattering functions at $T/J=0.015$
  and $0.0013$ in $\mathbf{Q}$-space: the blue lines bound the BZ
  while the red ones correspond to wavevector directions represented
  in panels (c,d).  (c,d) Scattering function at the same
  temperatures as function of energy and scattering vector for
  different directions in reciprocal space.  (e,f) Powder averaged
  dynamical scattering function versus energy and wave-vector modulus
  $|\mathbf{Q}|$.  (g,h) Constant energy cuts of the powder averaged
  scattering function versus wave-vector modulus $|\mathbf{Q}|$ for
  $\omega/J=0$ (red), 1 (blue), and 2 (green). Since the finite energy
  spectral weight is order of magnitudes weaker than the static one,
  the two latter energies have been multiplied by constant factors in
  order to superimpose all the constant energy cuts on the same plot.}
\label{fig:powder}
\end{figure*}

In conclusion, the numerical results show that the weathervane loop
fluctuations control the system relaxation. We identify two distinct
timescales associated with the inplane and out-of-plane fluctuations
and find that the temperature and wavevector dependences of these two
components are qualitatively consistent with loops diffusing in the
entropically induced potential well.
However, the exact role of incoherent thermal fluctuations remains
ambiguous and needs a better understanding. To go further, a thorough
numerical study in direct space (which is now in progress) is
required in order to separate more efficiently the dynamics of the
(local) loop motions from the other contributions.

\section{Comparison with experiments}
\label{sec:exp}

Experimental realizations of kagome antiferromagnet are often
complicated by further neighbor and/or anisotropic interactions, single
ion anisotropy, spin-lattice coupling, chemical imperfections as well
as lattice distortions
\cite{Simonet2008,Mutka2006,Schweika2007,Stewart2011,Coomer2006}. The
ground-state manifold is extremely unstable towards such
perturbations, which may partially or totally lift the degeneracy
\cite{Moessner2001}, so that any quantitative comparison with simple
models becomes difficult. The disappearance of the nematic order
parameter when the magnetic lattice contains defects, or the
stabilization of a $q=0$ ordered state when Dyaloshinski-Moryia
interactions are included, are two major illustrations of the effect of
perturbations \cite{Shender1993,Henley2001,Elhajal2002}. Nevertheless,
some compounds maintain a spin liquid behavior (often coexisting with
spin freezing) down to the lowest temperatures, which show qualitative
similarities with our present numerical work on the simple
antiferromagnetic Heisenberg model.

As described in sections \ref{sec:paraliquid} and \ref{ssec:coplanar},
fluctuations around the ground-state manifold show a complex
behavior which changes when the magnetic system tends towards
coplanarity.
In the liquid regime, spin relaxation is the result of incoherent
thermal fluctuations leading to an almost linear temperature
dependence of the relaxation rate. Such behavior was recently
observed by inelastic neutron scattering measurements in deuterium
jarosite, an experimental realization of a kagome lattice with spins
$S=5/2$ \cite{Fak2008}. The static correlations of this system are
very well reproduced by Monte Carlo simulations \cite{Fak2008}, so
 our classical approach could be fruitful to describe its
dynamical properties as well. Neutron measurements have been performed
from $T=14$ to 240~K, which, considering the effective coupling
constant $J_{cl}=JS^2=244$~K~\cite{Fak2008},
probes both the paramagnetic and liquid regimes ($0.05 \lesssim T/J
\lesssim 1$). The relaxation rates obtained experimentally and
numerically have the same order of magnitude and show qualitative
agreement over all the probed temperature range: for instance, data
collected at $T=240$~K (resp. $15$~K), giving $T/J=0.82$
(resp. $0.06$), provide a relaxation time $\tau J\simeq 3.2$
(resp. $10.5$), while $1.1(3)$ (resp. $6.2(6)$) is obtained from
numerical results.
Interestingly, a linear dependence of the fluctuation rate is also
observed in the quantum spin-$1/2$ kagome compound
Cu(1,3-benzenedicarboxylate) by means of muon spin spectroscopy
\cite{Marcipar2009}, with a relaxation time that is one order of
magnitude larger than predicted by the simulations. In this compound,
a saturation of the relaxation rate is observed at lower
temperatures. This could be due to the presence of sizable quantum
fluctuations which are not taken into account in the present study.

When approaching coplanarity, a distribution of time scales,
extending over approximately one order of magnitude for a given
temperature (see for instance Fig. \ref{fig:relax2}), is also observed
numerically. This distribution is induced by the entropic selection
that favors \s3s3 correlations and leads to a longer lifetime for this
type of spin configurations.  Such a time scale distribution has been
observed experimentally in the deuterium jarosite, for which a non
lorentzian line shape of the quasi-elastic intensity of the neutron
scattering data was observed at low temperatures \cite{Fak2008} .

Below the transition, the collective motion of the hexagonal loops
mostly control the spin relaxation in numerical studies. One
consequence is the apparition of a temperature independent second
timescale that is associated with the out-of-plane fluctuations of the
hexagonal loops. Recent experiments on Gadolinium Gallium Garnet
(GGG), a three dimensional generalization of the kagome lattice with
Heisenberg spins, also reported the observation of distinct
time-scales with very different temperature dependence. In this
system, the different time scales are associated with the simultaneous
development of short-range order dimerization dynamics, cooperative
paramagnetism, static order, and finally fluctuating ``protected" spin
clusters, so that the time scale distribution extends over several
order of magnitudes \cite{Bonville2004,Ghosh2008,Deen2010}. Dynamic
magnetization measurements also reported that the protected spin
clusters fluctuations are not thermally activated and do not depend on
temperature. They concluded that the protected spin clusters are
quantum dynamical objects\cite{Ghosh2008}. Our results suggest that
such a temperature independent behavior does not necessarily need
quantum fluctuations and may also be observed in classical
systems. However, for a more quantitative comparison, we should
consider the real three dimensional crystal structure of GGG as well
as the dipolar interactions, which have the same order of magnitude
than the exchange and whose role in the dynamical properties is still
unclear.

More generally, time-scale distributions are a feature of many
frustrated compounds, often characterized by the coexistence of a fast
dynamics together with a non-conventional glassy
behavior. Unfortunately, the glassy behaviors can not be observed
since the algorithm used for solving the dynamics does not accurately
describe long time dynamics. Freezing effects may however be studied
using stochastic spin dynamics method. Monte Carlo simulations applied
to the $q=3$ Potts model for instance show the presence of a freezing
time-scale, associated with the rearrangement of the clusters with a
typical length of few tens of spins\cite{Cepas2012}.

Finally, to complete this comparison, it is also necessary to discuss
the fastest spin dynamics $\propto J$, associated with spin wave
propagation.
Finite energy excitation exist in the two low temperature correlated
regimes\cite{Robert2008}, but theirs intensities are weak compared to
the quasistatic spectral weight. These spin waves-like excitation can
be identified in the scattering function $S(\mathbf{Q},t)$ as small
amplitude but high frequency modulations (of the order of
$\omega\simeq J$). However as shown on Fig.\ref{fig:powder} (c,d) for
two different directions in reciprocal space (see
Fig. \ref{fig:powder} (a)), spin waves excitation have clear signature
when we consider the scattering function $S(\mathbf{Q},\omega)$ in the
frequency domain.
The detailed analysis of these excitation along the high symmetry
direction $\mathbf{a}^\star$ $-$ where there is no quasi-static
spectral weight $-$ reveal that they are propagative in both coplanar
and cooperative regimes, although their lifetime $\tau_{SW} <
\tau_\perp \ll \tau_\parallel$ is strongly sensitive to the selection
of the coplanar ground state manifold \cite{Robert2008}.
It is therefore intriguing that no evident dispersive features have
been detected so far in liquid-like kagome compounds. In these
systems, single crystals are often not available because of technical
growing difficulties, so experiments are performed on powder
samples. The absence of dispersive excitation could then arise from
this powder averaging, which motivated us to calculate the excitation
spectrum for powder samples.

The powder averaged intensity maps in $(|\mathbf{Q}|,\omega)$ space
are shown in Fig. \ref{fig:powder} in the cooperative (e) and coplanar
(f) regimes. It appears that the inelastic excitation spectrum is
mostly dispersionless in both regimes in spite of existing propagative
spin waves in the single-crystal scattering function
(Fig. \ref{fig:powder} (c,d)). Indeed, the quasistatic fluctuations
(whose intensity is orders of magnitude larger than the spin wave
spectral weight) overwhelm the excitation spectrum and blur any
significant dispersive feature. Then, propagative effects may be very
difficult to observe experimentally on powder samples.

Constant energy cuts of the powder averaged scattering function,
displayed in Fig. \ref{fig:powder} (g,h) for $\omega=0$ (red), 1
(blue) and 2~$J$ (green), shows that an interval centered around the
energy $\omega/J=2$ should maximize the experimental detection of a
dispersive signal. Indeed, the $Q$-dependence of the scattering
function further evolves while approaching the top of the
dispersion. At this energy ($\omega/J=2$), the powder averaging gives
rise to a slightly more intense flat band in $|\mathbf{Q}|$. Its
intensity is smoothly structured with the scattering vector, and gives
broad maxima at different wave-vectors from the static scattering
function (see Fig. \ref{fig:powder} (g,h)). These results can be
compared to experimental results recently obtained in the volborthite,
a $S=1/2$ kagome compound which shows no signs of long-range order
down to 1.8~K in spite of an effective coupling of few tens of
kelvin. Although its static correlations and excitation spectrum
probably originates from a more complex exchange
Hamiltonian\cite{Nilsen2011,Yoshida2011} than the KHAFM, dispersive
excitation as well as a flat band at finite energy, likely resulting
from powder averaging, have been observed in inelastic neutron
scattering on powder samples \cite{Nilsen2011}.

\section{Conclusion}

The antiferromagnetic Heisenberg model on the kagome lattice is
blessed with very rich dynamics in all temperature regimes. Each
regime is characterized by a different mechanism of relaxation. At
high temperature, the relaxation of the magnetic phase is purely
diffusive.

When temperature reaches the cooperative regime, spins are still
disordered but algebraic spins correlations start to develop. They are
responsible for the exponential relaxation of the magnetic states at
short time scales with a relaxation time in $1/T$ in agreement with
previous studies. At long time scale, spin diffusion remains but it is
mediated by the dynamics of spins clusters rather than single spins as
in the paramagnetic regime.

In the very low temperature regime, entropic selection favors coplanar
states and a anisotropic dynamics. Although spin wave can propagate
through the system, their contribution to the relaxation is negligible
and limited to short times scales compared to the weathervane
defects. They are however very important for activating the
weathervane defects whose dynamics dominates the intermediate time
regime. A careful analysis of the relaxation shows that it is
anisotropic and depends on the direction of the fluctuations. The
characteristic times have a different temperature dependence, the
inplane component following an power law while the out-of-plane
component weakly depends on temperature. 

A more detailed study of the weathervane defects dynamics is needed to
understand the origin of the different temperature dependence of the
relaxation time observed in the lowest temperature regime.

\section*{Acknowledgements}

It is a great pleasure to acknowledge discussions with S. Viefers,
J. Chalker, O. Cepas, A. Ralko, P. Kopietz and S. Petit, as well as
N. Shannon and L. Jaubert for critical reading of the
manuscript. M. Taillefumier thanks the Max Planck Institute for the
complex systems for hosting him during the preparation of this
work. This work was performed on the Abel Cluster, owned by the
University of Oslo and the Norwegian metacenter for High Performance
Computing (NOTUR), and operated by the Department for Research
Computing at USIT, the University of Oslo
IT-department. M. Taillefumier acknowledges financial support from the
Norwegian research council and the University of Frankfurt.
\appendix

\section{Derivation of the dynamic structure factor for the $O(N)$ model}
\label{appendix:ncomp}

We describe in this appendix the derivation of the different
expressions given in the main section Sec.~\ref{sec:diffusion}. The
starting point of all calculations is the energy functional of the
$O(N)$ model
\begin{equation}
  \beta E = \frac12\sum_i \lambda s^2_i+\frac12 \beta J \sum_\alpha {\bf l}_\alpha^2,
  \label{eq:a1}
\end{equation}
where the index $\alpha$ represents the different triangles of the
kagome lattice and $l_\alpha$ is the sum of the components of the
spins forming the triangles. Eq.~(\ref{eq:a1}) differs from
Eq.~(\ref{eq:m:3}) by an additional term that is introduced to mimic the
behaviors the Heisenberg spins whose Lagrange multiplier $\lambda$ is
fixed by the condition $\left<s^2_i\right> = 1/3$.  Eq.\ref{eq:a1} can
conveniently be written as
\begin{equation}
  \label{eq:a2}
  \beta E = \frac12\sum_i \lambda s^2_i+ \beta J \sum_{i,j} {\bf s}_i
  (A^{\text{adj}}_{ij} +2 \delta_{ij}){\bf s}_j,
\end{equation}
where $A^{\text{adj}}_{ij}$ is the adjacent matrix of the kagome
lattice. By symmetry the adjacent matrix is
diagonal in ${\bf q}$ space so it is possible to express
Eq~\ref{eq:a2} in term of the collective variables ${\bf s}_i({\bf q})
= \sum_{\bf R} {\bf s}_{{\bf R},i} \exp (i {\bf q}(\cdot {\bf R}+{\bf
  r}_i)$ where the index $i$ is the sublattice index. The energy
functional is then given by
\begin{eqnarray}
  \label{eq:a3}
  \beta E &=& \frac12 \sum_i \lambda s_i^\dagger({\bf q}) s_i({\bf q})\nonumber \\ &+& \frac12 \beta J \sum_{ij}
  s^\dagger_i({\bf q}) (A^{\text{ad}}_{ij}({\bf q})+2\delta_{ij}) s_j({\bf
    q}).
\end{eqnarray}
$A^{\text{ad}}_{ij}({\bf q})$ are the matrix elements of the Fourier
transform of the adjacency matrix $A^{\text{ad}}({\bf q})$:
\begin{equation}
  \label{eq:a4}
  A^{\text{ad}}({\bf q}) = 2\left(\begin{array}{ccc}
      0 &\cos \frac{q_x}4 &\cos \frac{q_x+\sqrt{3}q_y}{4}\\
      \cos \frac{q_x}4 &0 &\cos\frac{q_x-\sqrt{3}q_y}{4}\\
      \cos\frac{q_x+\sqrt{3}q_y}{4} &\frac{q_x-\sqrt{3}q_y}{4} &0
    \end{array}\right).
\end{equation}
Then the eigenvalues of Eq.~(\ref{eq:l:2}) can be deduced from the
eigenvalues $\nu_l$ of $A^{\text{ad}}({\bf q})$ associated with the
eigenmodes ${\tilde s}_l({\bf q})$. We note $P({\bf q})$ the unitary
operator that transforms the operator (\ref{eq:a4}) in the diagonal
form. We find after some algebra that the eigenvalues of
$A^{\text{ad}}({\bf q})$ are given by
\begin{eqnarray}
  \label{eq:a5}
  \nu_1 &=& -2\\
  \nu_2 &=& 1 - \sqrt{3 + 2\cos q_x + 4\cos \frac{q_x}{2} \cos \frac{q_y
    \sqrt{3}}{2}}\\
\label{eq:a6}
  \nu_3 &=& 1 + \sqrt{3 + 2\cos q_x + 4\cos \frac{q_x}{2} \cos \frac{q_y
    \sqrt{3}}{2}}.
\label{eq:a7}
\end{eqnarray}

As explained in the main section, we describe the spin dynamics with a
Langevin equation given by Eq.\ref{eq:l:2}. The equation of motion of
the collective variables $s_{i}({\bf q})$ can be deduced by direct
calculation of the Fourier transform of Eq.~(\ref{eq:l:2}). We find
that
\begin{eqnarray}
  \label{eq:a8}
  \frac{d s_i({\bf q})}{dt} &=& \Gamma \left[(A^{\text{ad}}({\bf q})- z)\right.\nonumber\\
  &\times& \left.(T\lambda + J (A^{\text{ad}}({\bf q}) + 2)) [{\bf s}]({\bf q})\right]_i\nonumber\\ &+& \xi_i({\bf q},t),
\end{eqnarray}
where $\xi_i({\bf q},t) = \sum \xi_{i,r}(t) \exp(i {\bf q} \cdot {\bf
  r})$ is the Fourier transform of the white noise term $\xi_i(t)$ and
$[{\bf s}({\bf q})]$ is the vector formed by the collective variables
${\bf s}_i({\bf q})$. All indexes in Eq.~(\ref{eq:a5}) refer to the
sublattice index of the kagome net. After expressing Eq.~(\ref{eq:a8})
in the diagonal basis we find that
\begin{eqnarray}
  \label{eq:a9}
   \frac{d {\tilde s}_i({\bf q})}{dt} &=& \Gamma (\nu_i({\bf q})- z)(T\lambda + J (\nu_i({\bf q}) + 2)) {\tilde s}({\bf
      q})\nonumber\\ &+& \sum_jP^\dagger({\bf q})_{ij} \xi_j({\bf q},t),
\end{eqnarray}
where $P_{\alpha\beta}({\bf q})$ are the matrix elements of the operator
$P({\bf q})$. The solutions of Eq.~(\ref{eq:a9}) are given by
\begin{eqnarray}
  \label{eq:a10}
  {\tilde {\bf s}}_\alpha({\bf q}) &=& {\tilde {\bf s}}^0_\alpha({\bf q})\exp\left[-\frac{t}{\tau_\alpha}\right]\\ &+&
  \int_0^t P^\dagger_{i\alpha}({\bf q}) \xi_i({\bf q}, t^\prime)\exp\left[\frac{t^\prime-t}{\tau_\alpha}\right] dt^\prime
\end{eqnarray}
with
\begin{equation}
  \label{eq:a11}
  \tau_\alpha^{-1} = -\Gamma (\nu_\alpha-z)(T\lambda
  +J(\nu_\alpha+2)).
\end{equation}
using Eq.~(\ref{eq:a10}), we find that the spins correlations
functions are given by
\begin{equation}
  \label{eq:a12}
  \left<{\tilde {\bf s}_\alpha}({\bf q},t)|{\tilde {\bf s}_\beta}({\bf q},0)\right> = \frac{\delta_{\alpha\beta} T}{T\lambda+J(\nu_\alpha+2)}\exp\left[-\frac{t}{\tau_\alpha}\right].
\end{equation}
which combined with Eq.~(\ref{eq:m:6}) gives rise to
\begin{equation}
  \label{eq:a13}
  S({\bf q},t) = \sum_{\alpha} g_{\alpha}({\bf q})  \left<{\tilde {\bf s}_\alpha}({\bf q},t)|{\tilde {\bf s}_\alpha}({\bf q},0)\right>,
\end{equation}
and
\begin{equation}
  \label{eq:a14}
  g_{\alpha}({\bf q}) = \sum_{ij} P_{i\alpha}({\bf q})
  P_{j\alpha}({\bf q}).
\end{equation}
Using
\begin{equation}
  \label{eq:a15}
  \left<s_i ^ 2\right> = \frac{1}{3 N} \sum_{q \alpha} \frac{1}{\lambda + \beta \varepsilon_{\alpha}(q)} \approx \frac{1}{3\lambda} + O(T).
\end{equation}
we find that $\lambda = 1 + O(T)$ with $\left<s_i^2\right> = 1/3$ at low temperature.


\begin{thebibliography}{99}
\bibitem{Wannier1950}
  G. H. Wannier, Phys. Rev. {\bf 79}, 357 (1950). 
\bibitem{Kano1953}
  K. Kano and S. Naya, Prog. Theor. Phys. {\bf 10}, 158 (1953).
\bibitem{Toulouse1977} 
  G. Toulouse, Commun. Phys. {\bf 2}, 115 (1977)
\bibitem{Kar1979} See article of G. Toulouse in "Modern Trends in the
  Theory of Condensed Matter", A. Pekalski and J. Przystawa,
  editors. Springer-Verlag, 1979. Proceedings of XVI Karpacz Winter
  School of Theoretical Physics.
\bibitem{Villain1979}
  J. Villain, Z. Phys. B {\bf 33}, 31 (1979).

\bibitem{Kirk1977}
S. Kirkpatrick, \prb {\bf 16}, 46301¤741 (1977).
\bibitem{Moessner1998} R. Moessner and J. T. Chalker, \prl {\bf 80},
  2929 (1998), \prb {\bf 58}, 12049 (1998).
\bibitem{Reimers1993}
J. N. Reimers and A. J. Berlinsky, \prb {\bf 48}, 9539 (1993).
\bibitem{Reimers1992}
  J. N. Reimers, \prb {\bf 46}, 193 (1992).
\bibitem{Balents2010}
L. Balents, Nature {\bf 464}, 11 (2010).
\bibitem{Anderson1973}
  P. W. Anderson, Mat. Res. Bulletin {\bf 8}, 153 (1973).

\bibitem{Misguich2008}
  G. Misguich, Quantum spin liquids and fractionalization,  in Introduction to frustrated magnetism, ed. by C. Lacroix (Springer Heidelberg 2010).
\bibitem{Hermele2008}
M. Hermele, Y. Ran, P. A. Lee, and X. G. Wen, \prb {\bf 77}, 224413 (2008).
\bibitem{Harris1992}
  A. B. Harris, and C. Kallin, and A. L. Berlinsky, \prb {\bf 45}, 2899 (1992)
\bibitem{Gingras2011} M. J. P. Gingras, Spin Ices, Chapter 3,
  Introduction to Frustrated Magnetism, C. Lacroix, P. Mendels, and
  F. Mila, Springer Series in Solid-State Sciences (Springer, New
  York), Vol. 164. (2011).
\bibitem{Nisoli2013}
C. Nisoli, R. Moessner, and P. Schiffer, \rmp {\bf 85}, 1473 (2013).
\bibitem{Molavian2007} H. R. Molavian, M. J. P. Gingras and B. Canals,
  \prl {\bf 98}, 157204 (2007).
\bibitem{Harris1997}
M. J. Harris, S. T. Bramwell, D. F. McMorrow, T. Zeiske, and K. W. Godfrey,
\prl {\bf 79}, 2554 (1997).

\bibitem{Huse1992}
D. A. Huse and A. D. Rutenberg, \prb {\bf 45}, 7536 (1992).
\bibitem{Chalker1992} J. T. Chalker, P. C. W. Holdsworth, and
  E. F. Shender, \prl {\bf 68}, 855 (1992).
\bibitem{Sachdev1992}
  S. Sachdev, \prb {\bf 45},  12377 (1992).
\bibitem{Messio2012} L. Messio, B. Bernu, and C. Lhuillier,
  \prl {\bf 108}, 207204 (2012).
\bibitem{Yan2011}
S. Yan, D. A. Huse, and S. R. White, Science {\bf 332}, 1173 (2011).
\bibitem{Depenbrock2012} S. Depenbrock, I. P. McCulloch, and
  U. Schollw\"{o}ck, \prl {\bf 109}, 067201 (2012).

\bibitem{Iqbal2013}
Y. Iqbal, F. Becca, S. Sorella, and D. Poilblanc, \prb {\bf 87}, 060405(R) (2013).
\bibitem{Zhitomirsky2008}
M. E. Zhitomirsky, \prb {\bf 78}, 094423 (2008).
\bibitem{Henley2009}
C. L. Henley, \prb {\bf 80}, 180401 (2009).
\bibitem{Cepas2012}
O. C\'epas and B. Canals, \prb {\bf 86}, 024434 (2012).
\bibitem{Chern2013}
G.-W. Chern and R. Moessner, \prl {\bf 110}, 077201 (2013).

\bibitem{Keren1994}
A. Keren, \prl {\bf 72}, 3254 (1994).
\bibitem{Gardner2010}
J. S. Gardner, M. J. P. Gingras and J. E. Greedan, \rmp {\bf 68}, 53 (2010).
\bibitem{Conlon2009} P. H. Conlon and J. T. Chalker, \prl {\bf 102},
  237206 (2009), \prb {\bf 81}, 224413 (2010).
\bibitem{Robert2008}
J. Robert, V. Simonet, B. Canals and R. Ballou, \prl {\bf 101}, 117207 (2008).
\bibitem{Schnabel2012} S. Schnabel and D. P. Landau, \prb 86,
  014413 (2012).

\bibitem{Halperin1977}
  B. I. Halperin and W. M. Saslow, \prb {\bf 16}, 2154 (1977).
\bibitem{Ritchey1993}
I. Ritchey, P. Coleman and P. Chandra, Phys. Rev. Rapid Comm. B {\bf 47}, 15342 (1993).
\bibitem{Shender1993}
  E. F. Shender, V. B. Cherepanov, P. C. W. Holdsworth, and A. J. Berlinsky, \prl {\bf 70}, 3812 (1993).
\bibitem{Shender1995}
  E. F. Shender and P. C. W. Holdsworth J. Phys.: Condens. Matter {\bf 7}, 3295  (1995).
\bibitem{Sen2012}
  A. Sen, K. Damle, and R. Moessner, Phys Rev. B {\bf 86}, 205134 (2012).

\bibitem{Isakov2004}
  S. V. Isakov, K. Gregor, R. Moessner, and S. L. Sondhi, \prl {\bf 93}, 167204 (2004).
\bibitem{Henley2001}
  C. L. Henley,  Can. J. of Phys. {\bf 79}, 1307-1321 (2001).
\bibitem{Chandra1993}
  P. Chandra, P. Coleman and I. Ritchey, J. Phys. I France {\bf 3}, 591-610 (1993).
\bibitem{Hairer}
E. Hairer and G. Wanner, Solving Ordinary differential equations I and II, Springer (2000).
\bibitem{Creuz1987}
  M. Creutz, Phys. Rev. D {\bf 36}, 515 (1987).

\bibitem{Marshall1968}
  W. Marhsall abd R. D. Lowde, Rep.\ Prog.\ Phys. {\bf 31}, 705 (1968).
\bibitem{Halperin1969}
  B. I. Halperin and P. C. Hohenberg, Phys.\ Rev. {\bf 188}, 898 (1969).
\bibitem{Muller1988}
  G. M\"uller, \prl {\bf 60}, 2785 (1988), \prl {\bf 63}, 813 (1989)
\bibitem{Gerling1989} 
R. W. Gerling and D. P. Landau \prl {\bf 63}, 812 (1989), \prb {\bf 41}, 7139 (1990), \prb {\bf 42}, 8214 (1990).
\bibitem{Garanin1999}
  D. A. Garanin and B. Canals, \prb {\bf 59}, 443 (1999).
\bibitem{Henley2005}
  C. L. Henley, \prb {\bf 71}, 014424 (2005).

\bibitem{Lurie1974}
  N. A. Lurie, D. L. Huber and M. Blume, \prb {\bf 9}, 2171 (1974).
\bibitem{Huber2003}
  D. L. Huber, J.\ Phys.:\ Condens.\ Matter {\bf 15}, L579 (2003).
\bibitem{Mourigal2013}
M. Mourigal, W. T. Fuhrman, A. L. Chernyshev, and M. E. Zhitomirsky, \prb {\bf 88}, 094407 (2013).
\bibitem{Starykh2006}
Oleg A. Starykh, Andrey V. Chubukov, and Alexander G. Abanov, \prb {\bf 74}, 180403(R) (2006).
\bibitem{Henley1993}
  J. von Delft and C. L. Henley, \prb {\bf 48}, 965 (1993).
\bibitem{Cepas2011}
  O. Cepas and A. Ralko, \prb {\bf 84}, 020413(R) (2011).

\bibitem{Simonet2008}
  V. Simonet, R. Ballou, J. Robert, B. Canals, F. Hippert, P. Bordet, P. Lejay, P. Fouquet, J. Ollivier, and D. Braithwaite, \prl {\bf 100}, 237204 (2008).
\bibitem{Schweika2007}
  W. Schweika, M. Valldor, and P. Lemmens, \prl {\bf 98}, 067201 (2007).
\bibitem{Stewart2011}
  J. R. Stewart, G. Ehlers, H. Mutka, P. Fouquet, C. Payen, and R. Lortz, \prb {\bf 83}, 024405 (2011).
\bibitem{Mutka2006}
  H. Mutka, G. Ehlers, C. Payen, D. Bono, J. R. Stewart, P. Fouquet, P. Mendels, J. Y. Mevellec, N. Blanchard, and G. Collin, \prl {\bf 97}, 047203 (2006); M. Zbiri, H. Mutka, M. R. Johnson, H. Schober and Ch. Payen, \prb {\bf 81}, 104414 (2010).
\bibitem{Coomer2006}
  F. C. Coomer, A. Harrison, G. S. Oakley, J. Kulda, J. R. Stewart, J. A. Stride, B. Fak, J. W. Taylor and D. Visser,  J. Phys.: Condens. Matter {\bf 18}, 8847 (2006).

\bibitem{Moessner2001}
  R. Moessner, Can. J. of Phys. {\bf 79}, 1283 (2001).
\bibitem{Elhajal2002}
  M. Elhajal, B. Canals, and C. Lacroix, \prb {\bf 66}, 014422 (2002).
\bibitem{Fak2008}
  B. Fak, F. C. Coomer, A. Harrison, D. Visser and M. E. Zhitomirsky, Europhys.\ Lett. {\bf 81}, 17006 (2008).
\bibitem{Marcipar2009}
  L. Marcipar, O. Ofer, A. Keren, E. A. Nytko, D. G. Nocera, Y. S. Lee, J. S. Helton, and C. Bains, \prb {\bf 80}, 132402 (2009).
\bibitem{Ghosh2008}
  S. Ghosh, T. F. Rosenbaum, and G. Aeppli, \prl {\bf 101}, 157205 (2008).
\bibitem{Deen2010}
  P. P. Deen, O. A. Petrenko, G. Balakrishnan, B. D. Rainford, C. Ritter, L. Capogna, H. Mutka, and T. Fennell, \prb {\bf 82}, 174408 (2010).
\bibitem{Bonville2004}
  P. Bonville, J. A. Hodges, J. P. Sanchez, and P. Vulliet, \prl {\bf 92}, 167202 (2004).
\bibitem{Gotze2011} O. G\"otze, D. J. J. Farnell, R. F. Bishop,
  P. H. Y. Li, and J. Richter, \prb {\bf 84}, 224428 (2011).
\bibitem{Nilsen2011}
  G. J. Nilsen, F. C. Coomer, M. A. de Vries, J. R. Stewart, P. P. Deen, A. Harrison, and H. M. Ronnow, \prb {\bf 84}, 172401 (2011).
\bibitem{Yoshida2011}
  H. Yoshida, J. Yamaura, M. Isobe, Y. Okamoto, G. J. Nilsen and Z. Hiroi, Nature Comm. {\bf 3} 860 (2011).
\bibitem{HenleyReview}
  Christopher Henley, Ann. Rev. Condens. Matter Phys. {\bf 1}, 179 (2010).
\bibitem{Mohar2010}
  B. Mohar and J. Salas, J. Stat. Mech., P05016 (2010).
\end{thebibliography}
\end{document}